\documentclass[aps,pre,10pt,notitlepage,showpacs,nofootinbib,superscriptaddress,twocolumn,raggedbottom,preprintnumbers,longbibliography]{revtex4-1}

\usepackage{amsfonts}
\usepackage{multirow}
\usepackage{amsmath}
\usepackage{amssymb}
\usepackage{array}
\usepackage{bbold}
\usepackage{bm}
\usepackage{booktabs}
\usepackage{dsfont}
\usepackage{enumitem}
\usepackage[perpage]{footmisc}
\usepackage{array}
\usepackage{float}
\usepackage{graphicx}
\usepackage[utf8]{inputenc}
\usepackage{lastpage}
\usepackage{nicefrac}
\usepackage[normalem]{ulem}
\usepackage[dvipsnames]{xcolor}
\usepackage{graphicx}
\usepackage{color}
\usepackage{bigstrut}
\usepackage{enumitem}
\usepackage{bm,mwe}
\usepackage[export]{adjustbox}
\usepackage{listings}
\usepackage{nameref}
\usepackage{dcolumn}
\usepackage{textcomp}
\usepackage{adjustbox}
\usepackage{soul}

\usepackage{hyperref}
\hypersetup{
    colorlinks=true,
    urlcolor=blue,
    linkcolor = blue,
    urlcolor  = blue,
    citecolor = blue,
    anchorcolor = blue}

\graphicspath{{./plots/}}

\newcolumntype{R}[2]{%
    >{\adjustbox{angle=#1,lap=\width-(#2)}\bgroup}%
    l%
    <{\egroup}%
}

\begin{document}


\title{
Predicting the Mpemba Effect Using Machine Learning}

\author{Felipe Amorim}
\email{FelipeAugusto.deAmor@my.avemaria.edu}
\affiliation{Ave Maria University, Ave Maria, FL 34142, USA}
\author{Joey Wisely}
\email{Peter.Wisely@my.avemaria.edu}
\affiliation{Ave Maria University, Ave Maria, FL 34142, USA}

\author{Nathan Buckley}
\email{Nathan.Buckley@my.avemaria.edu}
\affiliation{Ave Maria University, Ave Maria, FL 34142, USA}

\author{Christiana DiNardo}
\email{Christiana.George@my.avemaria.edu}
\affiliation{Ave Maria University, Ave Maria, FL 34142, USA}

\author{Daniel Sadasivan}
\email{Daniel.Sadasivan@avemaria.edu}
\affiliation{Ave Maria University, Ave Maria, FL 34142, USA}

\preprint{}

\begin{abstract}
The Mpemba Effect  can be  studied with Markovian dynamics in a non-equilibrium thermodynamics framework. The Markovian Mpemba Effect can be observed in a variety of systems including the Ising model. We demonstrate that the Markovian Mpemba Effect can be predicted in the Ising model with several machine learning methods: the decision tree algorithm, neural networks, linear regression, and non-linear regression with the LASSO method. The positive and negative accuracy of these methods are compared. Additionally, we find that machine learning methods can be used to accurately extrapolate to data outside the range which they were trained. Neural Networks can even predict the existence of the Mpemba Effect when they are trained only on data in which the Mpemba Effect does not occur. This indicates that information about which coefficients result in the Mpemba Effect is contained in coefficients where the results does not occur. Furthermore, neural networks can predict that the Mpemba effect does not occur for positive $J$, corresponding to the ferromagnetic ising model even when they are only trained on negative $J$, corresponding to the anti-ferromagnetic ising model. All of these results demonstrate that the Mpemba Effect can be predicted in complex, computationally expensive systems, without explicit calculations of the eigenvectors.
 \end{abstract}

\maketitle


\section{Introduction}
\subsection{Mpemba Effect}
\label{sec:Intro/Mpemba}

The Mpemba Effect is best known for the claim that hot water can freeze faster than cool water in the same environment. This was named after Erasto Mpemba, who most famously brought it to the attention of the scientific community~\cite{Mpemba:1969} although it has been debated for thousands of years~\cite{aristotle-metaphysics-350BCE,nla.cat-vn251458,henry:2016}. However, the Mpemba Effect applies to a large number of situations beyond freezing water. We quote the definition given in Ref.~\cite{PhysRevX.9.021060} "The Mpemba effect is a counterintuitive relaxation phenomenon, where a system prepared at a hot temperature cools down faster than an identical system initiated at a cold temperature when both are quenched to an even colder bath." 
Several recent theoretical explanations of the effect have been demonstrated. One shows that the Mpemba Effect can be predicted from statistical mechanics~\cite{Lasanta:2017,Baity_Jesi_2019,Torrente_2019,Gij_n_2019,https://doi.org/10.48550/arxiv.2206.08846}. Another explanation, Refs.~\cite{lu:2017,PhysRevX.9.021060}
makes use of the framework developed in Ref.~\cite{lu:2017}, which considers the Markovian dynamics as a cooling process in the framework of non-equilibrium thermodynamics. We refer to this process as the Markovian Mpemba Effect (MME). The MME is further studied in Refs.~\cite{Busiello,PhysRevLett.131.017101} Ref.~\cite{lu:2017} studies the MME to an antiferromagnetic nearest-neighbor interacting Ising spin chain, generalizing the Mpemba Effect to other systems. The one-dimensional Ising spin chain will be the main model used in this paper, enabling us to use machine learning methods to predict the effect, as described in Sec.~\ref{sec:form}. 

The Mpemba Effect traditionally describes systems relaxing towards an equilibrium temperature that is lower than their initial temperature.  However, a reverse effect occurs when two systems relax towards equilibrium at a higher temperature. Sometimes the former is referred to as the Mpemba Effect and the latter is referred to as the Inverse Mpemba Effect, but both cases meet the definition of the Mpemba Effect used in this paper, namely that the system that is initially farther from equilibrium becomes closer  after a finite time. Unless otherwise specified we refer to the Inverse Markovian Mpemba Effect simply as the Mpemba Effect.

This paper is organized as follows. Sec.~\ref{subsec:appandim} discusses recent works with results that can be related to our results. 
Sec.~\ref{sec:form}, describe the formalism (laid out by Ref.~\cite{lu:2017,PhysRevX.9.021060}) used to determine whether  the Mpemba Effect occurs in the Ising model. Sec.~\ref{sec:datageneration} 
describes the data generated for machine learning and the methods used to generate this data. Sec.~\ref{sec:maclearn}, summarizes the various machine learning methods employed. The main results of this work are presented and discussed in Sec.~\ref{sec:results}. We conclude in Sec.~\ref{sec:conclusion}.

\subsection{Applications and Impact}
\label{subsec:appandim}

The Mpemba Effect has been observed in a number of systems. In addition to water,~\cite{Tang,jeng2006mpemba}, it has been predicted in magnetoresistance alloys~\cite{chaddah2010overtaking}, numerically predicted in colloids~\cite{Schwarzendahl}, predicted analytically and numerically in gasses~\cite{Megias,Biswas}, predicted in particles bounded by anharmonic potentials~\cite{Meibohm}, observed in computational simulations and experiments in  gas hydrates~\cite{ahn2016experimental}, and in computational simulations of carbon nanotube resonators~\cite{greaney2011mpemba}.  It has been demonstrated to have application for faster heating with precooling~\cite{PhysRevLett.124.060602}, that the effect happens for temperatures dependent on the maximum work that can be extracted from the system~\cite{Ch_trite_2021}, and  that the Mpemba Effect can lead to quantum heat engines with greater power output and stability~\cite{PhysRevE.105.014104}. 
Furthermore, the Mpemba Effect is closely linked to the Kovacs Memory Effect, by which systems out of equilibrium cannot merely be described by macroscopic thermodynamic variables. Knowledge of the Mpemba Effect can be applied when studying memory effects in Refs.~\cite{memoryeffect1,memoryeffect2,memoryeffect3,memoryeffect4,memoryeffect5}. Notably, it has been demonstrated that the phenomenon of eigenvalue crossing, closely linked to the Mpemba effect, can be understood as a first order phase transition~\cite{Teza2022-Eigenvalue}

These works demonstrate the potential applications of the effect and motivate better statistical understanding. To our knowledge, no work has used machine methods to predict the occurence of the ME in any system. 
The work presented in this paper uses learning methods to predict the Mpemba Effect in the Ising model. The comparison of methods done in this analysis could be applied to the prediction of the ME in other systems. 

Machine learning has previously been applied to the Ising model in a number of ways other than the Mpemba Effect: To extract the region between phases using autoencoding neural networks,~\cite{Walker_2020}, to predict probability distributions with neural networks~\cite{https://doi.org/10.48550/arxiv.1708.04622}, and comparison of various classification methods such as random forests, decision trees, k nearest neighbors and artificial neural networks for the prediction of magnetization~\cite{Portman_2017}. 

These works find several interesting results that can be compared to the work in this paper. Firstly, Ref.~\cite{https://doi.org/10.48550/arxiv.1708.04622} found that only the number of nodes in the first layer of the deep neural network could improve the accuracy of the prediction. Secondly, Ref.~\cite{Portman_2017} found that the Decision Tree method was the most accurate predictor. Even though these results predict different things, it is interesting to see whether these results, (namely that the decision tree algorithm is the most accurate predictor and that deep learning improves prediction) also apply to predicting the Mpemba effect in the Ising model.

\section{The Markovian Mpemba Effect Formalism}
\label{sec:form}

This work uses the formalism developed in Refs.~\cite{lu:2017,PhysRevX.9.021060} for the Mpemba Effect in the Ising model. We summarize the key equations in the following.

\begin{figure}[b]
\centerline{\includegraphics[trim=0 0 0 0,clip, scale=0.2]{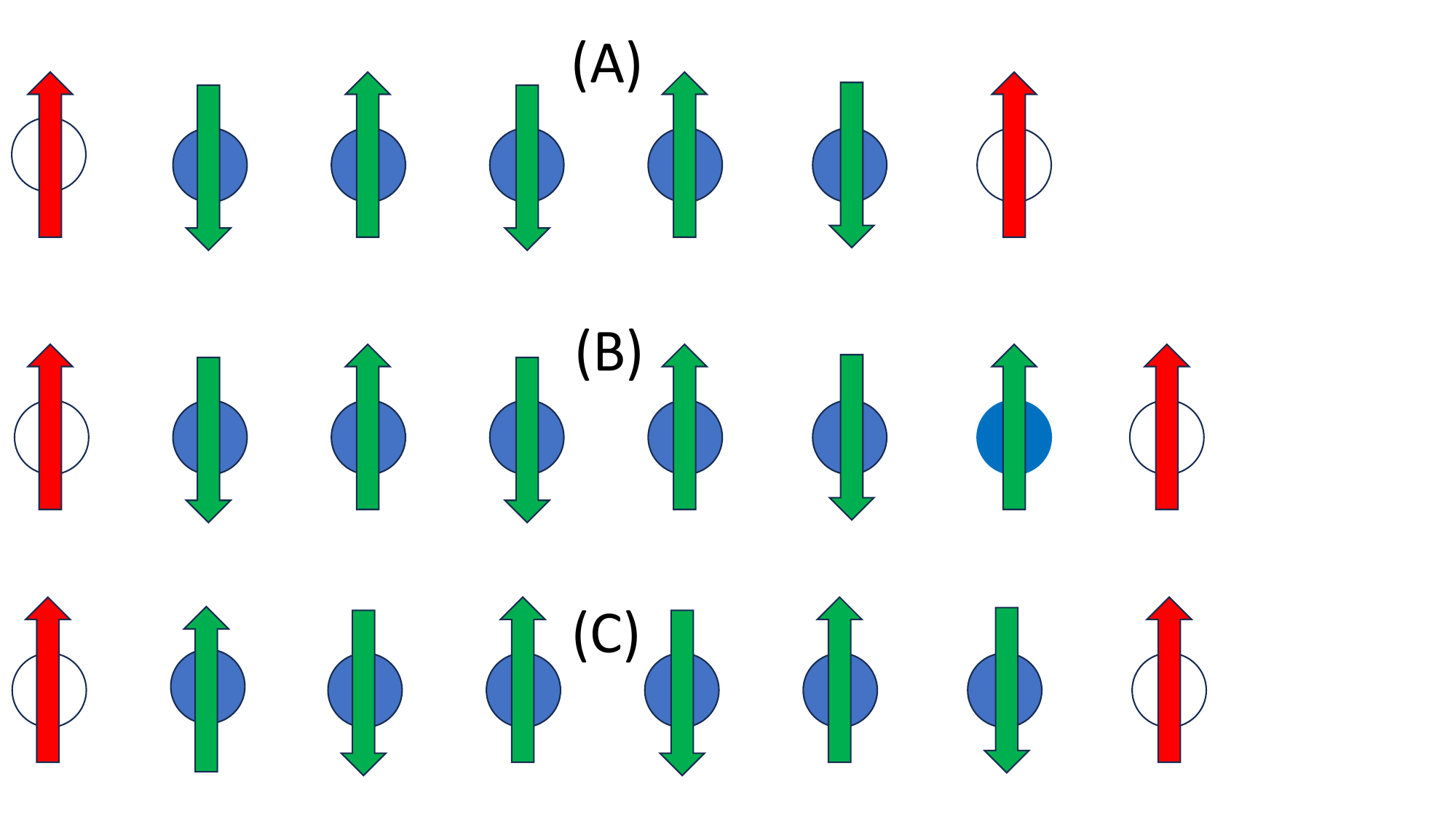}}
\caption{Ising chains of green arrows in blue shaded circles with fixed endpoints shown as red arrows in unshaded circles. For the anti-ferromagnetic Ising model, there is a non-degenerate ground state at 0 temperature where each spin is anti-aligned with its neighbors. This is shown for $N=5$ in the upper row (A). However, for even $N=6$, it is not possible for each spin to be anti aligned as shown in lower rows (B) and (C).  }
\label{fig:isingchain}
\end{figure}

The one-dimensional Ising model consists of a chain of $N$ spins, $s_i$ with values of 1 or -1. The energy for a given set of spins is defined to be 

\begin{align}
    E=-J\sum_{k=0}^N s_k s_{k+1}-h\sum_{k=1}^N s_k,
\label{eq:hamiltonian}
\end{align}
where $J$ and $h$ are parameters respectively giving the  field of interaction between neighboring spins and the strength of the external field. There are multiple ways to handle the end points, $s_1$ and $s_N$, of the Ising chain. One possibility is to consider both end points 
 to be connected to fixed spins as shown in Fig.~\ref{fig:isingchain} or Fig. 3 of Ref.~\cite{lu:2017}. This choice of endpoints mean that for odd $N$ the ground state energy level is non-degenerate whereas for even $N$, the ground state energy is degenerate. This can be seen in Fig.~\ref{fig:isingchain}

 Another possibility is to have the two endpoints connected such that the Ising chain forms a ring. Fig.1 of Ref.~\cite{Teza2021.125.110601} shows such a ring. 

A probability distribution $\vec{p}(t)$ that represents the probability of finding the system in one of the $2^N$ micro-states is defined for the case of a finite number of states, by the equation:
\begin{align}
    \frac{d \vec{p}(t)}{dt}= R(T_b)\vec{p}(t),
    \label{eq:diffyQ}
\end{align}
where $T_b$ is the bath temperature. Matrix multiplication is assumed. One approach is to use heat bath dynamics (employed in Refs.~\cite{lu:2017,PhysRevX.9.021060}). In this case, the transition rate matrix from a state $j$ to state $i$, $R_{ij}(T_b)$, is defined as:

\begin{align}
R_{ij} = \begin{cases}
\Gamma e^{-\frac{1}{2}\beta_b (E_{i} - E_j)},& \text{if i and j differ by one spin flip} \\
0,& \text{if i and j differ by} >1 \text{ spin flip} \\
-\sum _{k\neq j}R_{{kj}}, & i=j\\
\end{cases}\label{eq:Driving}
\end{align}

While this rate matrix can provide an exact description of the system's evolution over time starting from any initial probability distribution, the computational cost for larger $N$ are prohibitive. The method of coarse-graining can reduce the computational cost. In this method, microstates of the same energy are grouped together. This method is employed in Refs.~\cite{Teza_Thesis,Teza.125.110601,Teza2021.125.110601,Teza2022.2203.11644}. If coarse-graining is used, the rate matrix is given by Ref.~\cite{Teza2021.125.110601} (in the limit where the system is weakly coupled to the external environment) as
\begin{align}
R_{ij}^{weak}=\frac{G_{ij}}{\Omega_j} \frac{1}{1+e^{\beta_b (E_i-E_j)}, }
\label{eq:CoarseRateMatrix}
\end{align}
where $\Omega_{j}$ is the number of microstates with energy $E_i$ and $G_{ij}$ is the number of transitions between microstates of energy $E_i$ and $E_j$.
The coarse-graining approach can considerably reduce the size of the rate matrix. Instead of $2^N$ states, there are now $2+(N/2)^2$ states. This allows calculation for larger $N$. This work makes calculation for both methods. The former method with fixed endpoints is used in Secs.~\ref{subsec:ComparsionSmall}, \ref{subsec:Extrapolation}, and \ref{subsec:Strong}, and the latter method with the endpoints forming a ring is used in Sec.~\ref{subsec:ComparsionLarge}.

At a given temperature, a system will have an equilibrium state,

\begin{align}
\vec{\pi}(T)=\frac{e^{-E_i/k_BT}}{\sum_{i}e^{-E_i/k_BT}}.
\label{eq:Equilibrium}
\end{align}

Because the eigenvalues form a complete basis, any state can be expressed as,

\begin{align}
    \vec{p}=\vec{\pi}(T)+\sum_{i>1} a_i \vec{v}_i
\end{align}
where T is the temperature, $v_i$ are the eigenvectors of the rate transition matrix and $a_i$ are the coefficients corresponding to the vectors $v_i$. $\vec{\pi}$ is the eigenvector corresponding to an eigenvalue of 0. This eigenvector can be determined by the Boltzmann probability distribution. 
 
The general solution to Eq.~\eqref{eq:diffyQ} is given by

\begin{align}
\label{eq:ProbExpansion}
\vec{ p}(T;t) = e^{Rt}\vec{\pi}(T)= \vec{\pi}(T_b) + \sum _{i > 1} a_i(T) e^{\lambda_{i}t} \bm v_i,
\end{align} 

 and $\lambda_i$ are the eigenvalues of the transition rate matrix  and $T_b$ is the bath temperature.

The distance from equilibrium function, $D[\vec{p}(t); T_b]$, is given by

\begin{align}
    D[\vec{p}(t); T_b]=\sum_i \left( \frac{E_i (p_i-\pi_i^b)}{T_b}+p_i \ln{p_i}-\pi^b_i \ln{\pi_i^b} \right).
\end{align}

Multiple possible distance functions can be used. We use the distance function employed in Ref.~\cite{lu:2017}, however others are possible such as the Kullback-Leibler divergence used in Ref.~\cite{Teza2022-Eigenvalue}. The Mpemba Effect occurs for any two probability distributions, $\vec{p}_H$ and $\vec{p}_C$, if $D[\vec{p}_H(0); T_b]>D[\vec{p}_C(0); T_b]$ but $D[\vec{p}_H(t') T_b]<D[\vec{p}_C(t'); T_b]$ for some time, $t'$. Ref.~\cite{lu:2017} shows that this will occur if $|a_2^H|<|a_2^C|$. The $H$ in $\vec{p}_H$ refers to hot and the $C$ in $\vec{p}_C$ refers to cold. However, in the case of the Inverse Mpemba Effect in which the initial probability relaxes to a hotter bath temperature, the probability vector corresponding to the hotter initial system will have a smaller distance function than the vector corresponding to the colder system.

Additionally, Ref.~\cite{PhysRevX.9.021060} defines the Strong Mpemba Effect. This occurs when $a_2$ in Eq.~\eqref{eq:ProbExpansion} is 0. This means that, for large $t$, the time evolution is governed by $a_3$. A system with an initial $\vec{p}$ such that $a_2=0$ will approach equilibrium exponentially slower than a state with $a_2 \neq 0$. In contrast, cases where the Mpemba effect occurs when the hotter system has initial $\vec{p}$ with $a_2\neq0$ is known as the Weak Mpemba Effect. The Mpemba Effect can be further divided into the Direct Mpemba Efect in which the bath temperature is lower than the temperature of the initial states and the Inverse Mpemba Effect in which the bath temperature is lower. In this work, we primarily classify the Weak Inverse Mpemba Effect, however, we discuss classification of the Strong Inverse Mpemba Effect in Sec.~\ref{subsec:Strong}.

\section{Data Generation}
\label{sec:datageneration}

The data used for the various machine learning methods is generated in the following way. The matrix $R_{ij}$, given in Eq.~\eqref{eq:Driving}, is calculated for a given $J$, $h$, $T_b$, and $N$. Sparse Matrix methods are used to improve the speed of the computation. The Arnoldi method~\cite{arnoldi:principle} is used to calculate the second greatest (least negative) eigenvalue and the corresponding eigenvector. These correspond to $\lambda_2$ and $\bm{v}_2$ from Eq.~\eqref{eq:ProbExpansion}. These can be used to calculate $a_2$ using the method described of Sec. II in Ref.~\cite{PhysRevX.9.021060}.

 When we use smaller $N$ we use values between 5 and 15.  We also make predictions for data with larger $N$ in Sec.~\ref{subsec:ComparsionLarge}. The values of $h$ are between -10 and 10; the values of $J$ are between -10 and 0, and the values of $T$ are between 1 and 30.

For the data set with smaller $N$, we work only with odd-numbered spins. As discussed, the ground state with fixed endpoints for even $N$ is degenerate whereas it is non-degenerate for odd $N$. We find that the difference between the dynamics in the case of odd and even spins were large enough that none of the ML methods were able to accurately predict even $N$ in the case of small $N$ with fixed endpoints. We thus exclude even-valued $N$. This should be understood as a distinct limitation of the method. It should be noted that Monte-Carlo methods for numerically determining the Mpemba effect used  in Ref.~\cite{PhysRevLett.124.060602} do not suffer from this limitation. However, for larger $N$ in the weak coupling limit using Glauber dynamics, we are able to make accurate predictions for even $N$ even if the data are only trained on odd $N$.

For each data set, we work with a bath temperature of $T_b=30$. The bath temperature is a physical property of the system and could be varied for these methods. However, we work with a single bath temperature so that the temperatue of initially hotter system, $T_{h}$, and the temperature of the initially colder system, $T_{c}$, are selected from the same range. It is worth noting that, as far as the Mpembe effect is concerned, our results that do not vary $\beta_B$ but do vary $J$ can be mapped to results that vary $\beta_B$ but not $J$. This is because the time-evolution of a probability distribution (with an initial value given by  Eq.~\eqref{eq:Equilibrium}) depends on the rate matrix given by Eq.~\eqref{eq:Driving} which depends on the energy given in Eq.~\eqref{eq:hamiltonian}. The following mapping would leave $\vec{\pi(T)}$ and $R_{ij}$ unchanged:
\begin{align}
J\rightarrow \alpha J, \hspace{1.5 mm} h\rightarrow \alpha h, \hspace{1.5 mm} T_c\rightarrow \alpha T_c, \hspace{1.5 mm} 
T_h\rightarrow \alpha T_h, \hspace{1.5 mm} 
\beta_b\rightarrow  \beta_b/\alpha \hspace{5 mm} 
\end{align}
If $\alpha$ is chosen to be $\alpha=1/|J|$, then $J=-1$ for each data point and the new bath temperature becomes $\beta_b |J|$. Thus, such a mapping would allow results in this paper to be related to other results which vary $\beta_b$ but not $J$.

\section{Machine Learning Methods}
\label{sec:maclearn}

We use four different machine learning methods to predict the occurrence of the Mpemba Effect in an Ising chain.  All of the methods employed have certain free parameters called fit parameters and a loss function which measures the distance of the prediction from the data. The methods use various algorithms to fit the data or match the model to the data.

\subsection{Decision Tree Algorithm (DT)}
For the decision tree, the classification of data is divided into two classes: for each set of initial conditions, the system can either undergo the IMME or not.
 We use a data set  with in the format of:
\begin{equation}
\{N, J, h, T_{c}, T_{h} \}=    \begin{cases}
        1 & \text{if ME occurs } \\
        0 & \text{if ME does not occur},
    \end{cases}
    \label{eq:DirectData}
\end{equation}

The decision tree is trained using the CART (Classification And Regression Tree) algorithm~\cite{breiman:book84}. Just like all ML methods, this method minimizes a loss function. However, the loss function for this method is different than the loss function of the other methods used. Therefore, we can compare the accuracy of this method to the accuracy of other methods but cannot compare the loss function of this method to the other methods employed.

\subsection{Neural Network (NN)}

Neural networks can be trained to predict the parameters $|a_2|$ given a set of Mpemba parameters. Prediction of $a_2$ is sufficient to determine whether the Mpemba Effect occurs. The network is trained on 4 input parameters, \begin{equation}
\{N, J, h, T \}=   |a_2|
\end{equation}
Because all models rely of fit parameters, multiplying $|a_2|$ by a constant factor will not effect the accuracy. If the fit parameters of each model are multiplied by the same factor, the fitting procedure will be mathematically the same as if no factor had been applied to either. However, certain algorithms automatically select natural sized starting values. To ensure that the automatic starting values are optimized, we multiply $|a_2|$ by a factor of 10000. This leads to the exact same local minimum but allows for more efficient training. This adjustment prevents the need to change the starting parameters of each algorithm.

We work with a network consisting of 3 hidden layers with 50, 20, and 5 nodes respectively. Our explanation for this architecture is given in App.~\ref{sec:NeuralArchitecture}. Each layer as well as the output layer uses the Scaled Exponential Linear Unit (SELU) activation function. The loss function is the
Mean Squared Logarithmic Error (MSLE)~\cite{https://doi.org/10.1029/2021MS002681}. This is given with the formula 
\begin{align}
L(y,\hat{y})=\frac{1}{n}\sum_{i=0}^n\left( \log(y_i+1)-\log(\hat{y}+1)  \right)^2,
\label{eq:MSLE}
\end{align}
where  $y_i$ are the actual values, $n$ is the number of data points, and $\hat{y}_i$ are the predicted values.

The neural network can also be used to directly predict the Mpemba Effect rather than predicting $|a_2|$. This is done by training the network with data given by Eq.~\eqref{eq:DirectData}. The network remains the same except that the output layer has a sigmoid activation function and a binary cross-entropy loss function is used. The case where the Mpemba Effect is predicted directly is referred to as NN2.

\subsection{Linear Regression (LR)}
\label{subsec:LR}

Linear regression is used to predict the $|a_2|$ coefficients with the data. We once again minimize the MSLE given in Eq.~\eqref{eq:MSLE}. The predicted values, $\hat{y}_i$, are given by 
\begin{align}
\label{eq:linearregression}
 &\hat{y}_i=\sum_{j=0}^4 c_{j}x_{ij}. \\
\nonumber  x_{i0}=1; \hspace{2 mm} x_{i1}=N_i; &\hspace{2 mm} x_{i2}=J_i; \hspace{2 mm}  x_{i3}=h_i; \hspace{2 mm}  x_{i4}=T_i. 
\end{align} 
The fit parameters are $c_{j}$.

 The MSLE is chosen as the loss function rather than the sum of least squares because it more accurately captures the behavior of data with very different orders of magnitude.

Linear regression cannot be used to determine whether the  Mpemba 
Effect occurs because, for the effect to occur, $|a_2|$ must increase as $T$ increases. If $|a_2|$ is given by a monotonically decreasing linear function of $T$, the $T$ that is farther from equilibrium will always correspond to a larger value of $|a_2|$. 

Nevertheless, linear regression is useful for two reasons. Firstly, the MSLE can be compared to other methods, giving  information about their accuracy. Secondly, the values of the fit parameters $c_{j}$ can be used as a crude approximation of the importance of each category of data for predicting $|a_2|$. 

The fit parameters are given in Tab.~\ref{tab:correlations}. We also quote the correlation coefficents which correspond to an even simpler model. Note that because the linear regression in the fit was performed by minimizing the MSLE rather than the least squares, the regression coefficients are not the same as multivariate correlation coefficients.
\begin{table}
\begin{center}
\begin{tabular}{ |c||c|c|c|c|c| } 
\hline
&  $c_0$
 &$N$ &  $J$ & $h$ & $T$\\
\hline
\multirow{6}{0.0em}{}Regression &&&&&\\coefficients   & 10.191 & 84.359& -78.725 &  -9.889 & -36.903 \\ \hline
Correlation&&&&& \\coefficients & - & -0.005 & -0.370& -0.027 & -0.601 \\
\hline
\end{tabular}
\end{center}
\caption{\underline{Upper Row:} The linear regression fit parameters. The parameter in column $i$ corresponds to the fit parameters $c_i$ in Eq.~\eqref{eq:linearregression}. The first column is labeled $c_0$. The other columns are labeled with the parameters that $c_i$ is multiplied by. \underline{Lower Row:} The correlation coefficients between each parameter and $a_2$.} 
\label{tab:correlations}
\end{table}
\subsection{Nonlinear Regression (NLR) with the LASSO Method}

Non-linear models are able to have more fit parameters than linear models. We use a general expansion of the form

\begin{align}
\hat{y}_i= \sum_{k_1=0}^M \ldots 
\sum_{k_4=0}^M \delta_{M, \sum_i k_i} \scalebox{1.3}{c}_{\underbrace{1 \ldots}_{k_1 \text{ times}} \ldots \underbrace{4 \ldots}_{k_4 \text{ times}}} \prod_{j=1}^4
P_{k_j}(x_{ij})
\label{eq:lassofitfunction}
\end{align}
where $P_{k_j}(x_{ij})$ are the Legendre Polynomials mapped to the range of each data set, $M$ is the order up to which we consider, $\delta_{M, \sum_i k_i} $ is the Kronecker Delta Function, and $\scalebox{1.3}{c}_{\underbrace{1 \ldots}_{k_1 \text{ times}} \ldots \underbrace{4 \ldots}_{k_4 \text{ times}}}$ are the fit parameters. In the case when $M=1$, this expansion reduces to linear regression. The error minimized is once again the MSLE given in Eq.~\eqref{eq:MSLE}.
In our notation, coefficients such as $c_{1144}$ and $c_{4411}$ are multiplied by identical Legendre Polynomials. Each coefficient corresponds to an identical term as each other other coefficient with subscripts that are a permutation of its subscript. We include only one of each set of identical coefficients. 

The Legendre polynomials were chosen rather than simple polynomials because of their orthonormality. This allows for a continuous function to be fit with fewer free parameters than would be necessary with non-orthogonal functions. ~\cite{Legendre}. 

\begin{figure}[b]
\centerline{\includegraphics[scale=0.75]{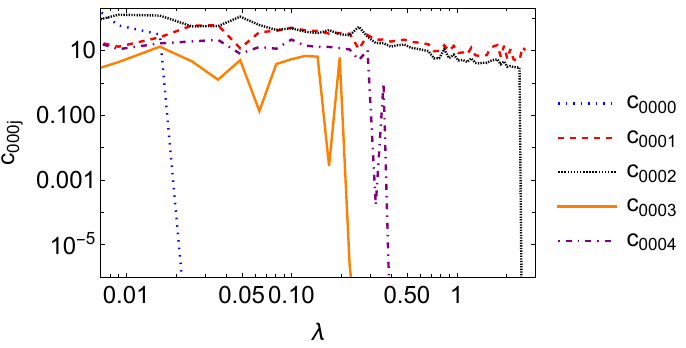}}
\caption{ The best fit values of parameters $c_{000j}$ for $j$ from 1 to 4 defined in Eq.~\eqref{eq:lassofitfunction}. The parameters are obtained by minimizing the loss function given in Eq.~\eqref{eq:LassoLossFunction} for various $\lambda$. As $\lambda$ increases, $c_{00ij}$ will become small. The order in which the parameters approach 0 gives the order of importance of each parameter in minimizing the MSLE. Note that the parameter $c_{0001}$ does not descend to 0 in this plot. If a larger range of $\lambda$ were chosen, it would.}
\label{fig:LassoParameters}
\end{figure}

For large $M$, the number of fit parameters lead to a prohibitive computational cost of the minimization. Not all fit parameters at a given order are equally effective at minimization. The LASSO Method \cite{10.2307/2346178,Guegan:2015mea,Sadasivan:2018jig,Landay:2018wgf} is employed in order to determine which fit parameters are most effective and which can be neglected to reduce computational cost. 
This is done in the following way. 
We randomly select 1000 data points and use them to calculate a modified loss function given by
\begin{align}
L(y,\hat{y})=\frac{1}{n}\sum_{i=0}^n\left( \log(y_i+1)-\log(\hat{y}+1)  \right)^2 +\lambda \sum|c_{ijk\ell} |.
\label{eq:LassoLossFunction}
\end{align}
The term $\lambda$ is a penalty term that can be varied. The data points are used to find the fit parameters that lead to local minima for various values of $\lambda$. When $\lambda=0$, all fit parameters are non-zero. As $\lambda$ becomes large, all fit parameters approach 0. A number of fits are performed with all fit parameters up to the fourth order with a gradually increasing $\lambda$. The fit parameters are ranked according to the value of $\lambda$ necessary to reach an absolute value less than $10^{-6}$. This list is ordered in terms of importance for the minimization of the MSLE. A visualization of the ordering can be found in Fig.~\ref{fig:LassoParameters} which shows the values of the second order parameters as $\lambda$ is increased. The lower the value of $\lambda$ at which the parameters are close to 0, the less important they are.

\begin{figure}[b]
\centerline{\includegraphics[scale=0.7]{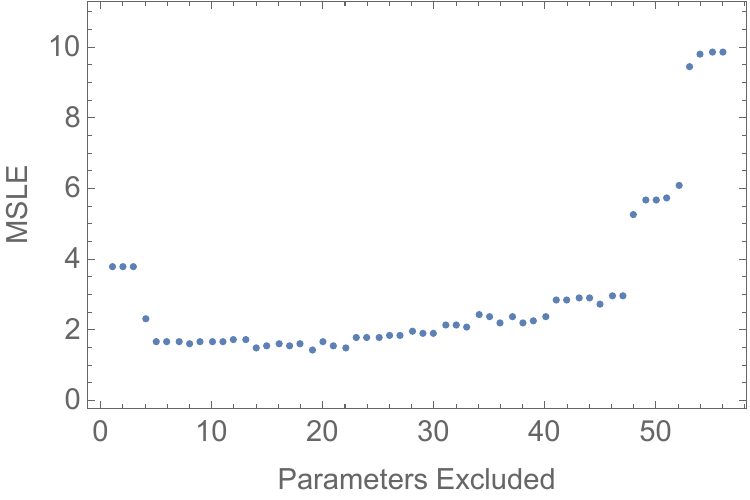}}
\caption{The validation MSLE varying the number of excluded parameters. The data were fit to 1000 training points. The parameters are excluded in order of impact as described in the text. The minimum corresponds to a number of excluded parameters that neither overfit nor underfit the 1000 data points. }
\label{fig:LassoFig}
\end{figure}

The parameters are then validated by calculating the MSLE with new data. First the data are validated with a fit involving all parameters. Next, the same validation is performed while excluding the parameter determined to be least important. After that, the least important two parameters are excluded. This process continues until all parameters have been excluded. This process generates a list of validation MSLE with parameters excluded in order of importance from least important to most important. The validation MSLE is plotted in Fig.~\ref{fig:LassoFig}.

The plot shows a minimum when 22 parameters are excluded. Thus 22 parameters are excluded for the non-linear regression. This allows for quicker minimization.

When fitting the data the following procedure is performed. Firstly, we fit the parameters $c_{000i}$ to the data from $i$ from 0 to $4$ using the Newton method. The other fit parameters are kept constant at 0. Next we fit the parameters $c_{00ij}$ with $i$ and $j$ from $0$ to $4$. The initial values of $c_{000i}$ are the best fit values from the previous minimization. The other fit parameters are given initial values 0. All parameters not used in the fit are once again set to 0. The same process is repeated for $c_{0ijk}$, and finally for $c_{ijk\ell}$.

This method of finding minimum parameters is employed in order to find a stable local minimum. If all  parameters were fitted at once without appropriate starting values, small changes in the fit such as changing the number of data points or increasing the penalty term would result in drastically different local minima. However, it causes the fit to rely most heavily on lower order parameters. 

The best fit parameters are given in Tab.~\ref{tab:lasso} in the Appendix. Perhaps because of the fit procedure, far more higher order parameters than lower order parameters are found by the LASSO to be unnecessary. The four fourth order parameters that are not excluded are $c_{1114}$, $c_{3334}$, and $c_{4441}$. All three of these parameters involve the temperature $T$. This could indicate that the correlations between the temperature and other input variables is important. This emphasis on the importance of $T$ in determining $a_2$ is in agreement with the linear correlation coefficients given in Tab.~\ref{tab:correlations}, however, as described in the next section, Non-Linear Regression is not found to be the most accurate method employed. Thus, any speculations  about the meaning of fourth order parameters is highly uncertain.


\section{Results and Discussion}
\label{sec:results}
\subsection{Comparison of Methods for small $N$}
\label{subsec:ComparsionSmall}
In this section, as well as Secs.~\ref{subsec:Extrapolation} and \ref{subsec:Strong}, we discuss the methods used to predict the Mpemba Effect for small $N$ (equal or less than 15). We use heat-bath dynamics with the rate matrix given in Eq.~\ref{eq:Driving} with endpoints connected to fixed up spins. Because both endpoints are connected to fixed spins, there is a considerable difference between the case when $N$ is odd and $N$ is even. For odd $N$, in the lowest energy state, each dipole will have the opposite spin of the two dipoles next to it. This is not possible for even $N$.  We were unable to obtain good predictions of the Mpemba effect for even $N$ in this section. This can be contrasted with the case of larger $N$ in a ring, in which we were able to obtain good predictions for even $N$ even when even values were not included in the training data.

\begin{figure}[b]
\centerline{\includegraphics[trim=0 0 0 0,clip, scale=0.65]{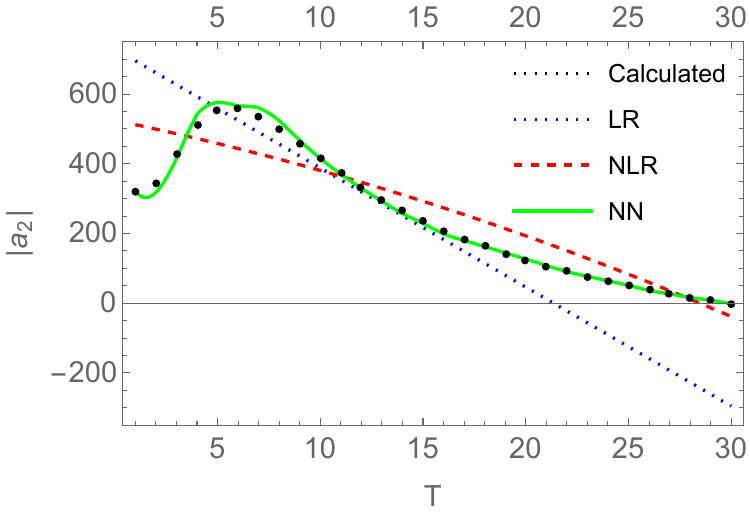}}
\caption{An example of $|a_2|$ as a function of temperature. The black dots show the actual calculations compared with the predictions for various methods (colored curves). This occurs for $N=7$, $J=-0.965$, $h=-7.01$. The NN gives the most accurate prediction followed by NLR and then LR. Note that the models are fit not to this data but to a large set of data including various $N$, $J$, $h$, and $T$.}
\label{fig:exampleMpemba}
\end{figure}

An example of the various methods compared with the data is shown in Fig.~\ref{fig:exampleMpemba}. Only the methods that predict $|a_2|$ directly are shown. The black dots give the actual computed data. At very low temperatures, they increase slightly with temperature indicating a very weak Mpemba Effect. After that, they decrease smoothly to equilibrium.

In order to compare the effectiveness of the various methods, several metrics are used. All of these metrics distinguish between training data, which are used to fix the free parameters or nodes of the model, and validation data, which are not used to determine the free parameters but are used to assess the accuracy. 

The validation MSLE can be used to compare the accuracy of the Neural Net (NN), Linear Regression (LR) and Non-Linear Regression (NLR). The Decision Tree (DT) and Neural Network trained directly on the Mpemba Effect (NN2) cannot be compared because they minimize different loss function.

The positive and negative validation accuracies are also computed. These are calculated by predicting whether the Mpemba Effect occurs for randomly selected $J$, $h$, $N$, and two temperatures. The positive accuracy is the percentage of cases where the method predicts the Mpemba Effect in which the effect actually occurs. The negative accuracy is the percentage of times the method correctly predicts that the effect does not occur. As discussed in Sec.~\ref{subsec:LR}, the accuracy is meaningless for LR.

For each method, the accuracy increases and the error decreases as the amount of data increases, however it eventually reaches a point at which additional data does not improve the predictions.
Tab.~\ref{tab:results} gives the calculations for the four methods for various numbers of data points.

The negative accuracy listed in the table for all four methods is much larger than the positive accuracy. This can be explained by the fact that the Mpemba Effect occurs in approximately $2\%$ of cases. Methods are more likely to minimize a loss function by predicting that the effect does not occur. For reference, a method that always predicts that the effect does not occur regardless of training data, would have a $0 \%$ positive accuracy and approximately $98 \%$ negative accuracy.  

This work establishes a baseline accuracy for predicting the Mpemba effect with various methods which can be used for future comparison of machine learning predictions. To compare the results in this work, we consider baseline "classifiers" of binary outcomes used in Refs~\cite{https://doi.org/10.48550/arxiv.1708.04622,XU2022100175,2023} to compare to predictions in a field never previously classified. The first of these is the Zero-Rate-Classifier (ZRC), which classifies all cases as the more likely of the outcomes. In this case, the more likely outcome is for the effect not to occur. Thus, the ZRC would have a total accuracy of 98$\%$. It is less helpful to use the ZRC to make a prediction of positive or negative accuracy. No matter what the data are, the ZRC will always give one outcome a 0$\%$. To establish a baseline for positive and negative accuracy, the Random Rate Classifier (RRC) is used. The RRC classifies by randomly selecting one of the binary with a likelihood proportional to the total fraction of positive or negative values. In this case, the RRC would guess that the effect that the effect does not occur with a likelihood of 98$\%$. This corresponds to a total accuracy of $0.98^2+0.02^2\approx 0.96$ and a positive and negative accuracy of $98\%$ and $2\%$ respectively. This is the baseline that our results can be compared to.

We find that for large data sets, NN2 is most accurate, though NN is almost as accurate and has the added benefit of predicting the values of $a_2$. 
For very small training sets, NLR is the most effective method at predicting $|a_2|$, although it fails to reach a high level of accuracy with larger data sets.

\begin{table}[h]
\centering
\begin{adjustbox}{width=\columnwidth,center}
\begin{tabular}{|r|r|r|r|r|r|}
    \hline
    Training & & & & &\\
    Points& DT & NN & NN2 & LR & NLR \\ \hline
    \multicolumn{6}{|c|}{\textbf{Error}} \\ \hline
    \multicolumn{1}{|l|}{100}  & - & 3.14 & - &4.86 & 2.44 
  \\ 
    \multicolumn{1}{|l|}{1000} &  - & 0.57 &  - & 4.08 & 1.65 \\ 
    \multicolumn{1}{|l|}{10000} &  -& 0.05 & - & 4.05& 1.77  \\ 
    \multicolumn{1}{|l|}{50000} 
 & -& 0.03 & - & 4.14 &  1.66  \\ 
    \multicolumn{1}{|l|}{200000} & -& 0.03 & - & 4.02 & 1.70  \\ \hline

    \multicolumn{6}{|c|}{\textbf{Positive Accuracy ($\%$)}} \\ \hline
    \multicolumn{1}{|l|}{100} & 1 $ \pm$ 14* & 13 $\pm$ 9.5 & 67 $\pm$ 18 &-&24 $\pm$ 21 
  \\ 
    \multicolumn{1}{|l|}{1000} &  30 $\pm$ 6 &18 $\pm$ 17 & 61 $\pm$ 14  &-&15 $\pm$ 8.5  \\ 
    \multicolumn{1}{|l|}{10000} & 46 $\pm$ 2.3 & 62 $\pm$ 8.5 & 79 $\pm$ 6.9 &-&17 $\pm$ 7.7 \\ 
    \multicolumn{1}{|l|}{50000} 
 & 64 $\pm$ 1.4 & 77 $\pm$ 5.9 & 79 $\pm$ 4.6  &-&20 $\pm$ 11   \\ 
    \multicolumn{1}{|l|}{200000}& 66 $\pm$ 2.3 & 82 $\pm$ 7.2  & 91 $\pm$ 5.2  &-&19 $\pm$ 00  \\ \hline

    \multicolumn{6}{|c|}{\textbf{Negative Accuracy} ($\%$)} \\ \hline
    \multicolumn{1}{|l|}{100}  &  98 $\pm$ 0.001 & 92 $\pm$ 4.5 * & 99 $\pm$ 0.6  &-&82 $\pm$ 13 *
  \\ 
    \multicolumn{1}{|l|}{1000} &  99 $\pm$ 0.5 &99 $\pm$ 0.9  & 99 $\pm$ 0.2  &-&89 $\pm$ 6.9 * \\ 
    \multicolumn{1}{|l|}{10000} & 99 $\pm$ 0.02 &99 $\pm$ 0.05 &  0.99  $\pm$ 0.09 &-&90 $\pm$ 7.6 *  \\ 
    \multicolumn{1}{|l|}{50000} 
 & 99 $\pm$ 0.02 & 99 $\pm$ 0.01  & 99 $\pm$ 0.03  &-&88 $\pm$ 4.5 *  \\ 
    \multicolumn{1}{|l|}{200000} & 99 $\pm$ 0.01 & 99 $\pm$ 0.01  & 99 $\pm$ 0.05  &-&88 $\pm$ 6.6 * \\ \hline

   \multicolumn{6}{|c|}{\textbf{Training Time} (seconds)} \\ \hline
    \multicolumn{1}{|l|}{100} & 0.58   & 0.4    &  0.8  & 0.2  &10
  \\ 
 \multicolumn{1}{|l|} {1000}   & 0.87   & 2.0   &   2.8 &  1.7 & 69
  \\ 
 \multicolumn{1}{|l|} {10000}  & 2.08   &  9.4  & 14  & 38.2  & 365
  \\ 
 \multicolumn{1}{|l|} {50000}  & 9.56   &  43   & 61   & 256   & 1624
  \\ 
\hline
     \multicolumn{6}{|c|}{\textbf{Computation Time} (seconds)} \\ \hline
    \multicolumn{1}{|l|}{50000}  & 1.41   & 1.52   &  1.48  & 0.20  & 1.71
  \\ 
\hline

\end{tabular}
\end{adjustbox}

\caption{The error, accuracy, and computational cost for several machine learning methods for varying numbers of training data points. The error is calculated with the MSLE on validation data. The positive/negative accuracy are the fraction of correct results when the method predicts that the effect occurs/does not occur. The * is placed next to numbers which do not beat the baseline RRC described in the text. The training time gives the time spent training the data and computation time gives the time for the machine learning method to validate a data set or make a prediction. Training time is not necessarily linear and thus we give the time for various sizes of training data sets. Computation time is linear so times for multiple data sets are not necessary. 
   }
    \label{tab:results}

\end{table}


\subsection{Extrapolation}
\label{subsec:Extrapolation}
In the previous section, the parameters for the validation data were different from those of the training data but they were generated in the same range. Machine learning can also be used to extrapolate by validating the model with data in a different range. This is particularly beneficial if the models can make accurate predictions on systems that are more computationally expensive than the systems they are trained on. Out of the parameters $N$, $J$, $h$, and $T$, only $N$ determines the computational complexity of the calculation. The time to compute $|a_2|$ does not depend on $J$, $h$ or $T$. 

\begin{figure}[b]
\centerline{\includegraphics[scale=0.6]{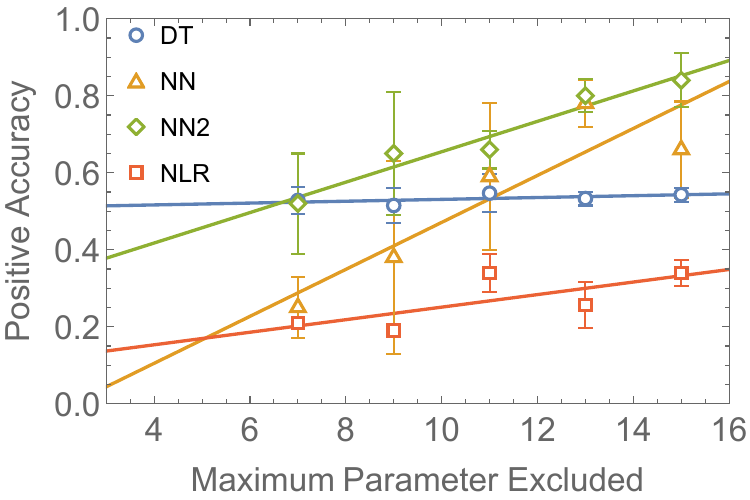}}
\caption{A plot of the positive accuracy vs. maximum value of $N$ excluded. The full data is given in Tab.~\ref{tab:extrapolation}. Best fit lines are shown to roughly estimate the relationship between the variables }
\label{fig:PositiveExtrapolation}
\end{figure}

Fig.~\ref{fig:PositiveExtrapolation} plots the positive accuracy of various methods for predicting the Mpemba Effect in the case when $N=15$, when it has only been trained on data with $N<15$. The full results are given in Tab.~\ref{tab:extrapolation}  in the Appendix. The lines shown in Fig.~\ref{fig:PositiveExtrapolation} give a rough approximation for how the accuracy depends on the maximum parameter excluded. Uncertainties are given by the standard deviation of 5 re-samples.

When values of $N$ close to 15 are included in the training data, the neural networks are once again the most accurate methods, however, the DT is relatively accurate at making predictions even when only $N=5$ is included. In fact, changing which values of $N$ are included has very little impact on the accuracy of the predictions, indicating that $N$ has very little effect in the prediction with this algorithm. These results agree with the crude prediction from the correlation coefficients, given in Tab.~\ref{tab:correlations}. 

It is reasonable to assume that the methods that are best at long range extrapolation in the ranges we consider will also be most effective at long range extrapolation outside the range we consider or for different systems than the Ising model. However, this has not been demonstrated for certain and merits further investigation.

Long range extrapolation could be useful due to the high computational cost for $a_2$ for higher $N$. The number of states for a given $N$ is $2^N$,  leading to a  $2^N\times2^N$ transition matrix. The computational cost of the eigenvalues for a $M\times M$ matrix is $\mathcal{O}(M^2)$~\cite{PresTeukVettFlan92}. Thus the computational cost for calculating $a_2$ is $2^{2N}$. Thus, a calculation with $N=5$ is quicker than a calculation with $N=15$ by a factor of $2^{20}$. Accurate prediction trained on much simpler systems could save considerable computational cost.
It should be noted that ML methods are not the only numerical methods to avoid computational costs. The Monte Carlo Methods used in Ref.~\cite{PhysRevLett.124.060602} are used to make calculations that would otherwise be impossible within realistic time constraints. For this method to be of maximum use, the computational cost recorded in Tab.~\ref{tab:results} would need to be lower than equivalent values for Monte Carlo methods.

\begin{table}[h]
\begin{center}
\begin{tabular}{ |c||c|c| } 
\hline
 &TME & ME \\
\hline
\multirow{3}{0.0em}{} Positive accuracy& 0.7764 ± 0.064 & 0.3035 ± 0.134\\ \hline
\multirow{3}{0.0em}{} Negative accuracy&0.8487 ± 0.070 & 0.9848 ± 0.012\\
\hline
\end{tabular}
\end{center}
\caption{The predictions for the Mpemba Effect when the neural network is only trained on data where the Mpemba Effect is not present. The neural network is trained on 20000 data points.
    }
\label{tab:mpembafromnompemba}
\end{table}

Notably, the Mpemba Effect can be predicted even if it is trained only with data in which it does not occur. Results for this extrapolation are given in Tab.~\ref{tab:mpembafromnompemba}.  The first prediction of the Mpemba Effect (ME) is the same as in the previous sections. The second method of prediction, referred to as the Total Mpemba Effect (TME), measures the accuracy of predicting whether or not the Mpemba Effect occurs for any temperature for a given $J$, $h$, and $N$. Put another way, neural networks can identify the non-monotonic temperature dependence of the coefficient $a_2$ even when they are only trained on monotonically decreasing $a_2$. This suggests that the Mpemba effect is encoded in the antiferromagnetic Ising model. This further confirms the results of Refs.~\cite{Teza2021.125.110601,Teza2022.2203.11644} 

The machine learning methods can also be extrapolated to $J$ outside the range the method was trained on. Specifically, our data set only includes negative $J$. No Mpemba Effect has been found for positive $J$. To test whether the neural networks can identify this, we made predictions on over $10^6$ data points for various $N$, and $h$. For each case, the algorithm correctly predicts that no Mpemba effect occurs for positive $J$. This result would not be as notable if the neural net were identifying that the effect never occurs for positive $J$ because more negative $J$ tend to cause the Mpemba effect. Fig.~\ref{fig:Jh-Plane} shows the regions where the Mpemba effect occurs on the $J-h$ Plane. The Mpemba effect does occur for negative values of $J$ close to 0 and does not occur for very negative values of $J$. This means that the neural network's identification that the effect only occurs for negative $J$ is a more subtle inference.

\subsection{The Strong Mpemba Effect}
\label{subsec:Strong}

In addition to the Mpemba Effect, machine learning methods can be used to predict the Strong Mpemba Effect, defined in Ref.~\cite{PhysRevX.9.021060}. In this situation, the hot system cools exponentially faster than the cold system. 

The Strong Mpemba effect occurs when $a_2=0$ and $T\neq T_b$. The neural network is trained on three input variables, $N$, $J$, and $h$. The output is 1 if the strong Mpemba Effect occurs for any temperature; the output is 0 if the Strong Mpemba Effect does not occur. Results are given in Tab.~\ref{tab:StrongMpemba}. 
\begin{table}[h]
\begin{tabular}{ 
|p{2.8cm}||p{2.8cm}|}
\hline
Positive accuracy & Negative Accuracy \\
\hline
0.4633 $\pm$ 0.043 & 0.995 $\pm$ 0.0016\\
\hline
\end{tabular}
\caption{The validation accuracy for the Strong Mpemba Effect. It is trained with 20000 data points.
    }
\label{tab:StrongMpemba}
\end{table}

\begin{figure}[h]
\centerline{\includegraphics[trim=0 0 0 0,clip, scale=0.6]{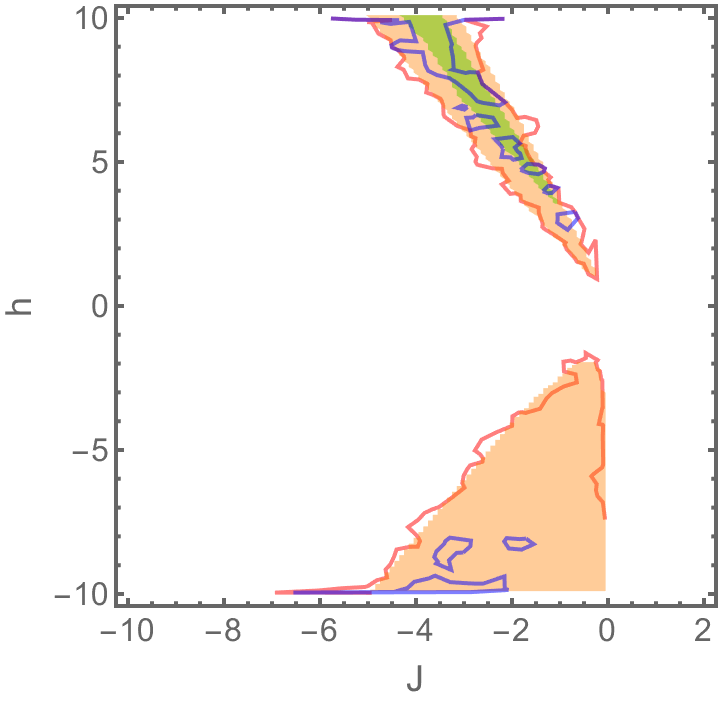}}
\caption{The $J$-$h$ plane. The shaded orange region (lighter shaded region) shows the NN2 prediction for the region in which the Mpemba effect occurs. The red contour (lighter contour) shows the region generated with a contour from the data. The shaded green region (darker shaded region) shows the predicted region for the strong Mpemba effect and the blue contour (darker contour) shows the contour generated from the data. We observe good agreement between the results and the prediction, especially for the ordinary Mpemba Effect.}
\label{fig:Jh-Plane}
\end{figure}

The machine learning algorithm can be used to generate plots of the $J$-$h$ Plane. Fig.~\ref{fig:Jh-Plane} shows a plot of which values the Mpemba effect occurs at and which values the strong Mpemba effect occurs at. This is compared to the predictions made by the neural net. This diagram is inspired by the phase diagrams of Ref.~\cite{Teza2021.125.110601} (Fig. 2) although it differs in that it plot the $J$-$h$ plane rather than the $T_b$-$h$ plane.

\subsection{Predictions for Large $N$ using Coarse Graining}
\label{subsec:ComparsionLarge}

Much larger values of $N$ can be computed for a given comptuational cost using the method of coarse-graining. We generate a data set using the rate matrix given in Eq.~\eqref{eq:CoarseRateMatrix}. This data set includes $a_2$ for the weak coupling limit for spins in a ring rather than with fixed end-points.

To test the methods, we generate a data set with odd numbered $N$ from 15 to $53$ as well as sets with $N=59$ and $N=50$. This range is comparable to the size of the Ising chain in Ref.~\cite{Teza2021.125.110601}, however an exact coarse-graining allowed a computation of an Ising chain with 1000 spins in Ref.~\cite{Teza2022.2203.11644}. Furthermore, The thermodynamic limit of large $N$ has been studied through the phenomenon of eigenvalues crossing (closely linked to the Mpemba Effect)~\cite{Teza2022-Eigenvalue}.

We find that NN1 and NN2 can be used to make accurate predictions even outside the range that the model is trained on. Results are shown in Tab.~\ref{tab:coarse}.

Notably, predictions can be made for $N=50$. As noted, in the previous sections, predictions could not be made for even $N$, however, with larger $N$ that form a ring, predictions can be made for even $N$ even is only odd $N$ are included in the training data. However, the positive accuracy of each of these methods noticeably decreases when more values of $N$ are excluded. This suggests a limitation to long-ranged extrapolation for these ising chains.

\begin{table}[h]
\begin{center}
\begin{tabular}{ |c|c|c|c|c| c| c| } 
\hline
 Excluded & Predicted & NN1+ & NN2+   & NN1- & NN2-   &Error \\
\hline

55, 57, 59 & 59 &  0.401 &  0.438 & 0.972 & 0.935 & 14.2 \\

 53 & 53 & 0.731 & 0.715 & 0.981 & 0.988 & 3.47  \\

  51, 53 & 53 & 0.530& 0.627 & 0.932 & 0.988 & 5.91 \\

    49, 51, 53 & 53 & 0.475 & 0.517 & 0.986 & 0.989 & 8.64 \\
50 & 50 & 0.515 & 0.985& 0.974  & 0.838 & 3.31 \\
\hline
\hline 

\hline - & All & 0.972 & 0.946& 0.781 & 0.760 & 4.12 \\
    - & 53 & 0.991 & 0.986 & 0.669  & 0.799 &  5.69\\
    \hline

    \hline
\end{tabular}
\label{tab:coarse}
\end{center}
\caption{ Validation accuracy and error of predictions by NN1 and NN2. Error is for NN1 only. Extrapolations are given above the double line. In this case, all values of $N$ listed as excluded in addition to all values of $N$ greater than the predicted value are excluded.  Below the double line, predictions for validation data within the range of the training data are included.
    }
\end{table}

\section{Conclusion}
\label{sec:conclusion}

The Mpemba Effect has been shown to occur in many systems, beyond the freezing of water. These applications motivate better predictions of the effect. In order to understand this effect, this work applies statistical methods to the Mpemba Effect in the Ising model.

For the case of 5-15 spins in an ising chain, we demonstrate that a number of machine learning methods can be used. Neural networks are the most effective method when a large enough training data set is used. For very small data sets, non-liner regression with the LASSO method may be more effective at predicting $a_2$.  These methods can predict the Mpemba Effect in systems much more complex than those on which they were trained. The decision tree method may be the most effective method at making predictions far outside the range it was trained on. Additionally, the Mpemba Effect can be predicted with neural networks when it is trained only on data where the Mpemba Effect does not occur. This indicates that information about the Mpemba Effect exists in situations in which it does not occur. Furthermore, we demonstrate that the strong Mpemba Effect can also be predicted using machine learning methods. It should be noted that a limitation to these methods (in comparision to numerical methods in other works) is that they can only be successfully applied to chains with an odd number of spins.

In addition, we have studied the effect for a larger number of spins in the spin-chain using coarse-graining in the weak-coupling limit. In this method, neural networks can be used to make accurate prediction for the Mpemba Effect. For these spin-chains, accurate predictions can be made for an even number of spins, even if the network is only trained with data from odd-numbered spin chains. However, a limitation of predicting the Mpemba Effect in these systems is that long-ranged extrapolation to much larger spin-chains than the networks are trained on is not as accurate.

This work can be extended in several ways. Firstly, future analysis could vary the bath temperature. Such an investigation could more optimally test the effect by selecting a single bath temperature and scanning for non-monoticity. Futhermore, these predictions were performed for the Mpemba Effect in the Ising model, however, the relative accuracy of each method might hold for the Mpemba Effect in other systems. Determination of the accuracy of various methods in other systems merits further investigation.
  

\bigskip

\textit{Acknowledgments} ---
This work was funded by an Ave Maria University undergraduate research grant by Michael and Lisa Schwartz. Thanks to Tom\'as Licheri for completing crucial tasks necessary for this research.

\bigskip

\bibliography{BIB}

\begin{thebibliography}{49}%
\makeatletter
\providecommand \@ifxundefined [1]{%
 \@ifx{#1\undefined}
}%
\providecommand \@ifnum [1]{%
 \ifnum #1\expandafter \@firstoftwo
 \else \expandafter \@secondoftwo
 \fi
}%
\providecommand \@ifx [1]{%
 \ifx #1\expandafter \@firstoftwo
 \else \expandafter \@secondoftwo
 \fi
}%
\providecommand \natexlab [1]{#1}%
\providecommand \enquote  [1]{``#1''}%
\providecommand \bibnamefont  [1]{#1}%
\providecommand \bibfnamefont [1]{#1}%
\providecommand \citenamefont [1]{#1}%
\providecommand \href@noop [0]{\@secondoftwo}%
\providecommand \href [0]{\begingroup \@sanitize@url \@href}%
\providecommand \@href[1]{\@@startlink{#1}\@@href}%
\providecommand \@@href[1]{\endgroup#1\@@endlink}%
\providecommand \@sanitize@url [0]{\catcode `\\12\catcode `\$12\catcode
  `\&12\catcode `\#12\catcode `\^12\catcode `\_12\catcode `\%12\relax}%
\providecommand \@@startlink[1]{}%
\providecommand \@@endlink[0]{}%
\providecommand \url  [0]{\begingroup\@sanitize@url \@url }%
\providecommand \@url [1]{\endgroup\@href {#1}{\urlprefix }}%
\providecommand \urlprefix  [0]{URL }%
\providecommand \Eprint [0]{\href }%
\providecommand \doibase [0]{http://dx.doi.org/}%
\providecommand \selectlanguage [0]{\@gobble}%
\providecommand \bibinfo  [0]{\@secondoftwo}%
\providecommand \bibfield  [0]{\@secondoftwo}%
\providecommand \translation [1]{[#1]}%
\providecommand \BibitemOpen [0]{}%
\providecommand \bibitemStop [0]{}%
\providecommand \bibitemNoStop [0]{.\EOS\space}%
\providecommand \EOS [0]{\spacefactor3000\relax}%
\providecommand \BibitemShut  [1]{\csname bibitem#1\endcsname}%
\let\auto@bib@innerbib\@empty
\bibitem [{\citenamefont {Mpemba}\ and\ \citenamefont
  {Osborne}(1969)}]{Mpemba:1969}%
  \BibitemOpen
  \bibfield  {author} {\bibinfo {author} {\bibfnamefont {E~B}\ \bibnamefont
  {Mpemba}}\ and\ \bibinfo {author} {\bibfnamefont {D~G}\ \bibnamefont
  {Osborne}},\ }\bibfield  {title} {\enquote {\bibinfo {title} {Cool?}}\ }\href
  {\doibase 10.1088/0031-9120/4/3/312} {\bibfield  {journal} {\bibinfo
  {journal} {Physics Education}\ }\textbf {\bibinfo {volume} {4}},\ \bibinfo
  {pages} {172--175} (\bibinfo {year} {1969})}\BibitemShut {NoStop}%
\bibitem [{\citenamefont {Aristotle}(350BCE)}]{aristotle-metaphysics-350BCE}%
  \BibitemOpen
  \bibfield  {author} {\bibinfo {author} {\bibnamefont {Aristotle}},\ }\href
  {http://classics.mit.edu/Aristotle/metaphysics.html} {\emph {\bibinfo {title}
  {Metaphysics}}},\ edited by\ \bibinfo {editor} {\bibfnamefont {Translated}\
  \bibnamefont {by~W.~D.~Ross}}\ (\bibinfo  {publisher} {The Internet Classics
  Archive},\ \bibinfo {year} {350BCE})\BibitemShut {NoStop}%
\bibitem [{\citenamefont {Bacon}(1677)}]{nla.cat-vn251458}%
  \BibitemOpen
  \bibfield  {author} {\bibinfo {author} {\bibfnamefont {Francis}\ \bibnamefont
  {Bacon}},\ }\href@noop {} {\emph {\bibinfo {title} {The Novum Organum of Sir
  Francis Bacon}}}\ (\bibinfo  {publisher} {Printed for Thomas Lee London},\
  \bibinfo {year} {1677})\ pp.\ \bibinfo {pages} {[3], 32 p.}\BibitemShut
  {Stop}%
\bibitem [{\citenamefont {Burridge}\ and\ \citenamefont
  {Linden}(2016)}]{henry:2016}%
  \BibitemOpen
  \bibfield  {author} {\bibinfo {author} {\bibfnamefont {Henry~C.}\
  \bibnamefont {Burridge}}\ and\ \bibinfo {author} {\bibfnamefont {F.~Paul}\
  \bibnamefont {Linden}},\ }\bibfield  {title} {\enquote {\bibinfo {title}
  {Questioning the mpemba effect: hot water does not cool more quickly than
  cold},}\ }\href {\doibase https://doi.org/10.1038/srep37665} {\bibfield
  {journal} {\bibinfo  {journal} {Sci. Rep.}\ }\textbf {\bibinfo {volume} {6}}
  (\bibinfo {year} {2016}),\ https://doi.org/10.1038/srep37665}\BibitemShut
  {NoStop}%
\bibitem [{\citenamefont {Klich}\ \emph {et~al.}(2019)\citenamefont {Klich},
  \citenamefont {Raz}, \citenamefont {Hirschberg},\ and\ \citenamefont
  {Vucelja}}]{PhysRevX.9.021060}%
  \BibitemOpen
  \bibfield  {author} {\bibinfo {author} {\bibfnamefont {Israel}\ \bibnamefont
  {Klich}}, \bibinfo {author} {\bibfnamefont {Oren}\ \bibnamefont {Raz}},
  \bibinfo {author} {\bibfnamefont {Ori}\ \bibnamefont {Hirschberg}}, \ and\
  \bibinfo {author} {\bibfnamefont {Marija}\ \bibnamefont {Vucelja}},\
  }\bibfield  {title} {\enquote {\bibinfo {title} {Mpemba index and anomalous
  relaxation},}\ }\href {\doibase 10.1103/PhysRevX.9.021060} {\bibfield
  {journal} {\bibinfo  {journal} {Phys. Rev. X}\ }\textbf {\bibinfo {volume}
  {9}},\ \bibinfo {pages} {021060} (\bibinfo {year} {2019})}\BibitemShut
  {NoStop}%
\bibitem [{\citenamefont {Lasanta}\ \emph {et~al.}(2017)\citenamefont
  {Lasanta}, \citenamefont {Vega~Reyes}, \citenamefont {Prados},\ and\
  \citenamefont {Santos}}]{Lasanta:2017}%
  \BibitemOpen
  \bibfield  {author} {\bibinfo {author} {\bibfnamefont {Antonio}\ \bibnamefont
  {Lasanta}}, \bibinfo {author} {\bibfnamefont {Francisco}\ \bibnamefont
  {Vega~Reyes}}, \bibinfo {author} {\bibfnamefont {Antonio}\ \bibnamefont
  {Prados}}, \ and\ \bibinfo {author} {\bibfnamefont {Andr\'es}\ \bibnamefont
  {Santos}},\ }\bibfield  {title} {\enquote {\bibinfo {title} {When the hotter
  cools more quickly: Mpemba effect in granular fluids},}\ }\href {\doibase
  10.1103/PhysRevLett.119.148001} {\bibfield  {journal} {\bibinfo  {journal}
  {Phys. Rev. Lett.}\ }\textbf {\bibinfo {volume} {119}},\ \bibinfo {pages}
  {148001} (\bibinfo {year} {2017})}\BibitemShut {NoStop}%
\bibitem [{\citenamefont {Baity-Jesi}\ \emph {et~al.}(2019)\citenamefont
  {Baity-Jesi}, \citenamefont {Calore}, \citenamefont {Cruz}, \citenamefont
  {Fernandez}, \citenamefont {Gil-Narvi{\'{o}}n}, \citenamefont
  {Gordillo-Guerrero}, \citenamefont {I{\~{n}}iguez}, \citenamefont {Lasanta},
  \citenamefont {Maiorano}, \citenamefont {Marinari}, \citenamefont
  {Martin-Mayor}, \citenamefont {Moreno-Gordo}, \citenamefont {Sudupe},
  \citenamefont {Navarro}, \citenamefont {Parisi}, \citenamefont
  {Perez-Gaviro}, \citenamefont {Ricci-Tersenghi}, \citenamefont
  {Ruiz-Lorenzo}, \citenamefont {Schifano}, \citenamefont {Seoane},
  \citenamefont {Taranc{\'{o}}n}, \citenamefont {Tripiccione},\ and\
  \citenamefont {Yllanes}}]{Baity_Jesi_2019}%
  \BibitemOpen
  \bibfield  {author} {\bibinfo {author} {\bibfnamefont {Marco}\ \bibnamefont
  {Baity-Jesi}}, \bibinfo {author} {\bibfnamefont {Enrico}\ \bibnamefont
  {Calore}}, \bibinfo {author} {\bibfnamefont {Andres}\ \bibnamefont {Cruz}},
  \bibinfo {author} {\bibfnamefont {Luis~Antonio}\ \bibnamefont {Fernandez}},
  \bibinfo {author} {\bibfnamefont {Jos{\'{e} }~Miguel}\ \bibnamefont
  {Gil-Narvi{\'{o}}n}}, \bibinfo {author} {\bibfnamefont {Antonio}\
  \bibnamefont {Gordillo-Guerrero}}, \bibinfo {author} {\bibfnamefont {David}\
  \bibnamefont {I{\~{n}}iguez}}, \bibinfo {author} {\bibfnamefont {Antonio}\
  \bibnamefont {Lasanta}}, \bibinfo {author} {\bibfnamefont {Andrea}\
  \bibnamefont {Maiorano}}, \bibinfo {author} {\bibfnamefont {Enzo}\
  \bibnamefont {Marinari}}, \bibinfo {author} {\bibfnamefont {Victor}\
  \bibnamefont {Martin-Mayor}}, \bibinfo {author} {\bibfnamefont {Javier}\
  \bibnamefont {Moreno-Gordo}}, \bibinfo {author} {\bibfnamefont
  {Antonio~Mu{\~{n}}oz}\ \bibnamefont {Sudupe}}, \bibinfo {author}
  {\bibfnamefont {Denis}\ \bibnamefont {Navarro}}, \bibinfo {author}
  {\bibfnamefont {Giorgio}\ \bibnamefont {Parisi}}, \bibinfo {author}
  {\bibfnamefont {Sergio}\ \bibnamefont {Perez-Gaviro}}, \bibinfo {author}
  {\bibfnamefont {Federico}\ \bibnamefont {Ricci-Tersenghi}}, \bibinfo {author}
  {\bibfnamefont {Juan~Jesus}\ \bibnamefont {Ruiz-Lorenzo}}, \bibinfo {author}
  {\bibfnamefont {Sebastiano~Fabio}\ \bibnamefont {Schifano}}, \bibinfo
  {author} {\bibfnamefont {Beatriz}\ \bibnamefont {Seoane}}, \bibinfo {author}
  {\bibfnamefont {Alfonso}\ \bibnamefont {Taranc{\'{o}}n}}, \bibinfo {author}
  {\bibfnamefont {Raffaele}\ \bibnamefont {Tripiccione}}, \ and\ \bibinfo
  {author} {\bibfnamefont {David}\ \bibnamefont {Yllanes}},\ }\bibfield
  {title} {\enquote {\bibinfo {title} {The mpemba effect in spin glasses is a
  persistent memory effect},}\ }\href {\doibase 10.1073/pnas.1819803116}
  {\bibfield  {journal} {\bibinfo  {journal} {Proceedings of the National
  Academy of Sciences}\ }\textbf {\bibinfo {volume} {116}},\ \bibinfo {pages}
  {15350--15355} (\bibinfo {year} {2019})}\BibitemShut {NoStop}%
\bibitem [{\citenamefont {Torrente}\ \emph {et~al.}(2019)\citenamefont
  {Torrente}, \citenamefont {L{\'{o} }pez-Casta{\~{n}}o}, \citenamefont
  {Lasanta}, \citenamefont {Reyes}, \citenamefont {Prados},\ and\ \citenamefont
  {Santos}}]{Torrente_2019}%
  \BibitemOpen
  \bibfield  {author} {\bibinfo {author} {\bibfnamefont {Aurora}\ \bibnamefont
  {Torrente}}, \bibinfo {author} {\bibfnamefont {Miguel~A.}\ \bibnamefont
  {L{\'{o} }pez-Casta{\~{n}}o}}, \bibinfo {author} {\bibfnamefont {Antonio}\
  \bibnamefont {Lasanta}}, \bibinfo {author} {\bibfnamefont {Francisco~Vega}\
  \bibnamefont {Reyes}}, \bibinfo {author} {\bibfnamefont {Antonio}\
  \bibnamefont {Prados}}, \ and\ \bibinfo {author} {\bibfnamefont
  {Andr{\'{e}}s}\ \bibnamefont {Santos}},\ }\bibfield  {title} {\enquote
  {\bibinfo {title} {Large mpemba-like effect in a gas of inelastic rough hard
  spheres},}\ }\href {\doibase 10.1103/physreve.99.060901} {\bibfield
  {journal} {\bibinfo  {journal} {Physical Review E}\ }\textbf {\bibinfo
  {volume} {99}} (\bibinfo {year} {2019}),\
  10.1103/physreve.99.060901}\BibitemShut {NoStop}%
\bibitem [{\citenamefont {Gij{\'{o} }n}\ \emph {et~al.}(2019)\citenamefont
  {Gij{\'{o} }n}, \citenamefont {Lasanta},\ and\ \citenamefont
  {Hern{\'{a}}ndez}}]{Gij_n_2019}%
  \BibitemOpen
  \bibfield  {author} {\bibinfo {author} {\bibfnamefont {A.}~\bibnamefont
  {Gij{\'{o} }n}}, \bibinfo {author} {\bibfnamefont {A.}~\bibnamefont
  {Lasanta}}, \ and\ \bibinfo {author} {\bibfnamefont {E.~R.}\ \bibnamefont
  {Hern{\'{a}}ndez}},\ }\bibfield  {title} {\enquote {\bibinfo {title} {Paths
  towards equilibrium in molecular systems: The case of water},}\ }\href
  {\doibase 10.1103/physreve.100.032103} {\bibfield  {journal} {\bibinfo
  {journal} {Physical Review E}\ }\textbf {\bibinfo {volume} {100}} (\bibinfo
  {year} {2019}),\ 10.1103/physreve.100.032103}\BibitemShut {NoStop}%
\bibitem [{\citenamefont {Meg{\'{\i}}as}\ and\ \citenamefont
  {Santos}(2022)}]{https://doi.org/10.48550/arxiv.2206.08846}%
  \BibitemOpen
  \bibfield  {author} {\bibinfo {author} {\bibfnamefont {Alberto}\ \bibnamefont
  {Meg{\'{\i}}as}}\ and\ \bibinfo {author} {\bibfnamefont {Andr{\'{e}}s}\
  \bibnamefont {Santos}},\ }\bibfield  {title} {\enquote {\bibinfo {title}
  {Mpemba-like effect protocol for granular gases of inelastic and rough hard
  disks},}\ }\href {\doibase 10.3389/fphy.2022.971671} {\bibfield  {journal}
  {\bibinfo  {journal} {Frontiers in Physics}\ }\textbf {\bibinfo {volume}
  {10}} (\bibinfo {year} {2022}),\ 10.3389/fphy.2022.971671}\BibitemShut
  {NoStop}%
\bibitem [{\citenamefont {Lu}\ and\ \citenamefont {Raz}(2017)}]{lu:2017}%
  \BibitemOpen
  \bibfield  {author} {\bibinfo {author} {\bibfnamefont {Zhiyue}\ \bibnamefont
  {Lu}}\ and\ \bibinfo {author} {\bibfnamefont {Oren}\ \bibnamefont {Raz}},\
  }\bibfield  {title} {\enquote {\bibinfo {title} {Nonequilibrium
  thermodynamics of the markovian mpemba effect and its inverse},}\ }\href
  {\doibase 10.1073/pnas.1701264114} {\bibfield  {journal} {\bibinfo  {journal}
  {Proceedings of the National Academy of Sciences}\ }\textbf {\bibinfo
  {volume} {114}},\ \bibinfo {pages} {5083--5088} (\bibinfo {year}
  {2017})}\BibitemShut {NoStop}%
\bibitem [{\citenamefont {Busiello}\ \emph {et~al.}(2021)\citenamefont
  {Busiello}, \citenamefont {Gupta},\ and\ \citenamefont {Maritan}}]{Busiello}%
  \BibitemOpen
  \bibfield  {author} {\bibinfo {author} {\bibfnamefont {Daniel}\ \bibnamefont
  {Busiello}}, \bibinfo {author} {\bibfnamefont {Deepak}\ \bibnamefont
  {Gupta}}, \ and\ \bibinfo {author} {\bibfnamefont {Amos}\ \bibnamefont
  {Maritan}},\ }\bibfield  {title} {\enquote {\bibinfo {title} {Inducing and
  optimizing markovian mpemba effect with stochastic reset},}\ }\href {\doibase
  10.1088/1367-2630/ac2922} {\bibfield  {journal} {\bibinfo  {journal} {New
  Journal of Physics}\ }\textbf {\bibinfo {volume} {23}} (\bibinfo {year}
  {2021}),\ 10.1088/1367-2630/ac2922}\BibitemShut {NoStop}%
\bibitem [{\citenamefont {Teza}\ \emph
  {et~al.}(2023{\natexlab{a}})\citenamefont {Teza}, \citenamefont {Yaacoby},\
  and\ \citenamefont {Raz}}]{PhysRevLett.131.017101}%
  \BibitemOpen
  \bibfield  {author} {\bibinfo {author} {\bibfnamefont {Gianluca}\
  \bibnamefont {Teza}}, \bibinfo {author} {\bibfnamefont {Ran}\ \bibnamefont
  {Yaacoby}}, \ and\ \bibinfo {author} {\bibfnamefont {Oren}\ \bibnamefont
  {Raz}},\ }\bibfield  {title} {\enquote {\bibinfo {title} {Relaxation
  shortcuts through boundary coupling},}\ }\href {\doibase
  10.1103/PhysRevLett.131.017101} {\bibfield  {journal} {\bibinfo  {journal}
  {Phys. Rev. Lett.}\ }\textbf {\bibinfo {volume} {131}},\ \bibinfo {pages}
  {017101} (\bibinfo {year} {2023}{\natexlab{a}})}\BibitemShut {NoStop}%
\bibitem [{\citenamefont {Tang}\ \emph {et~al.}(2022)\citenamefont {Tang},
  \citenamefont {Huang}, \citenamefont {Zhang}, \citenamefont {Liu},\ and\
  \citenamefont {Zhao}}]{Tang}%
  \BibitemOpen
  \bibfield  {author} {\bibinfo {author} {\bibfnamefont {Zhiqiang}\
  \bibnamefont {Tang}}, \bibinfo {author} {\bibfnamefont {Weidong}\
  \bibnamefont {Huang}}, \bibinfo {author} {\bibfnamefont {Yagang}\
  \bibnamefont {Zhang}}, \bibinfo {author} {\bibfnamefont {Yanxia}\
  \bibnamefont {Liu}}, \ and\ \bibinfo {author} {\bibfnamefont {Lin}\
  \bibnamefont {Zhao}},\ }\bibfield  {title} {\enquote {\bibinfo {title}
  {Direct observation of the mpemba effect with water: Probe the mysterious
  heat transfer},}\ }\href {\doibase 10.1002/inf2.12352} {\bibfield  {journal}
  {\bibinfo  {journal} {InfoMat}\ } (\bibinfo {year} {2022}),\
  10.1002/inf2.12352}\BibitemShut {NoStop}%
\bibitem [{\citenamefont {Jeng}(2006)}]{jeng2006mpemba}%
  \BibitemOpen
  \bibfield  {author} {\bibinfo {author} {\bibfnamefont {Monwhea}\ \bibnamefont
  {Jeng}},\ }\bibfield  {title} {\enquote {\bibinfo {title} {The mpemba effect:
  When can hot water freeze faster than cold?}}\ }\href
  {https://doi.org/10.1119/1.2186331} {\bibfield  {journal} {\bibinfo
  {journal} {American Journal of Physics}\ }\textbf {\bibinfo {volume} {74}},\
  \bibinfo {pages} {514--522} (\bibinfo {year} {2006})}\BibitemShut {NoStop}%
\bibitem [{\citenamefont {Chaddah}\ \emph {et~al.}(2010)\citenamefont
  {Chaddah}, \citenamefont {Dash}, \citenamefont {Kumar},\ and\ \citenamefont
  {Banerjee}}]{chaddah2010overtaking}%
  \BibitemOpen
  \bibfield  {author} {\bibinfo {author} {\bibfnamefont {P}~\bibnamefont
  {Chaddah}}, \bibinfo {author} {\bibfnamefont {S}~\bibnamefont {Dash}},
  \bibinfo {author} {\bibfnamefont {Kranti}\ \bibnamefont {Kumar}}, \ and\
  \bibinfo {author} {\bibfnamefont {A}~\bibnamefont {Banerjee}},\ }\bibfield
  {title} {\enquote {\bibinfo {title} {Overtaking while approaching
  equilibrium},}\ }\href@noop {} {\bibfield  {journal} {\bibinfo  {journal}
  {arXiv preprint arXiv:1011.3598}\ } (\bibinfo {year} {2010})}\BibitemShut
  {NoStop}%
\bibitem [{\citenamefont {Schwarzendahl}\ and\ \citenamefont
  {Löwen}(2022)}]{Schwarzendahl}%
  \BibitemOpen
  \bibfield  {author} {\bibinfo {author} {\bibfnamefont {Fabian~Jan}\
  \bibnamefont {Schwarzendahl}}\ and\ \bibinfo {author} {\bibfnamefont
  {Hartmut}\ \bibnamefont {Löwen}},\ }\bibfield  {title} {\enquote {\bibinfo
  {title} {Anomalous cooling and overcooling of active colloids},}\ }\href
  {\doibase 10.1103/physrevlett.129.138002} {\bibfield  {journal} {\bibinfo
  {journal} {Physical Review Letters}\ }\textbf {\bibinfo {volume} {129}}
  (\bibinfo {year} {2022}),\ 10.1103/physrevlett.129.138002}\BibitemShut
  {NoStop}%
\bibitem [{\citenamefont {Megías}\ \emph {et~al.}(2022)\citenamefont
  {Megías}, \citenamefont {Santos},\ and\ \citenamefont {Prados}}]{Megias}%
  \BibitemOpen
  \bibfield  {author} {\bibinfo {author} {\bibfnamefont {Alberto}\ \bibnamefont
  {Megías}}, \bibinfo {author} {\bibfnamefont {Andres}\ \bibnamefont
  {Santos}}, \ and\ \bibinfo {author} {\bibfnamefont {Antonio}\ \bibnamefont
  {Prados}},\ }\bibfield  {title} {\enquote {\bibinfo {title} {Thermal versus
  entropic mpemba effect in molecular gases with nonlinear drag},}\ }\href
  {\doibase 10.1103/PhysRevE.105.054140} {\bibfield  {journal} {\bibinfo
  {journal} {Physical Review E}\ }\textbf {\bibinfo {volume} {105}} (\bibinfo
  {year} {2022}),\ 10.1103/PhysRevE.105.054140}\BibitemShut {NoStop}%
\bibitem [{\citenamefont {Biswas}\ \emph {et~al.}(2022)\citenamefont {Biswas},
  \citenamefont {v~v},\ and\ \citenamefont {Rajesh}}]{Biswas}%
  \BibitemOpen
  \bibfield  {author} {\bibinfo {author} {\bibfnamefont {Apurba}\ \bibnamefont
  {Biswas}}, \bibinfo {author} {\bibfnamefont {Prasad}\ \bibnamefont {v~v}}, \
  and\ \bibinfo {author} {\bibfnamefont {R.}~\bibnamefont {Rajesh}},\
  }\bibfield  {title} {\enquote {\bibinfo {title} {Mpemba effect in
  anisotropically driven inelastic maxwell gases},}\ }\href {\doibase
  10.1007/s10955-022-02891-w} {\bibfield  {journal} {\bibinfo  {journal}
  {Journal of Statistical Physics}\ }\textbf {\bibinfo {volume} {186}}
  (\bibinfo {year} {2022}),\ 10.1007/s10955-022-02891-w}\BibitemShut {NoStop}%
\bibitem [{\citenamefont {Meibohm}\ \emph {et~al.}(2021)\citenamefont
  {Meibohm}, \citenamefont {Forastiere}, \citenamefont {Adeleke-Larodo},\ and\
  \citenamefont {Proesmans}}]{Meibohm}%
  \BibitemOpen
  \bibfield  {author} {\bibinfo {author} {\bibfnamefont {Jan}\ \bibnamefont
  {Meibohm}}, \bibinfo {author} {\bibfnamefont {Danilo}\ \bibnamefont
  {Forastiere}}, \bibinfo {author} {\bibfnamefont {Tunrayo}\ \bibnamefont
  {Adeleke-Larodo}}, \ and\ \bibinfo {author} {\bibfnamefont {Karel}\
  \bibnamefont {Proesmans}},\ }\bibfield  {title} {\enquote {\bibinfo {title}
  {Relaxation-speed crossover in anharmonic potentials},}\ }\href {\doibase
  10.1103/PhysRevE.104.L032105} {\bibfield  {journal} {\bibinfo  {journal}
  {Physical Review E}\ }\textbf {\bibinfo {volume} {104}} (\bibinfo {year}
  {2021}),\ 10.1103/PhysRevE.104.L032105}\BibitemShut {NoStop}%
\bibitem [{\citenamefont {Ahn}\ \emph {et~al.}(2016)\citenamefont {Ahn},
  \citenamefont {Kang}, \citenamefont {Koh},\ and\ \citenamefont
  {Lee}}]{ahn2016experimental}%
  \BibitemOpen
  \bibfield  {author} {\bibinfo {author} {\bibfnamefont {Yun-Ho}\ \bibnamefont
  {Ahn}}, \bibinfo {author} {\bibfnamefont {Hyery}\ \bibnamefont {Kang}},
  \bibinfo {author} {\bibfnamefont {Dong-Yeun}\ \bibnamefont {Koh}}, \ and\
  \bibinfo {author} {\bibfnamefont {Huen}\ \bibnamefont {Lee}},\ }\bibfield
  {title} {\enquote {\bibinfo {title} {Experimental verifications of
  mpemba-like behaviors of clathrate hydrates},}\ }\href@noop {} {\bibfield
  {journal} {\bibinfo  {journal} {Korean Journal of Chemical Engineering}\
  }\textbf {\bibinfo {volume} {33}},\ \bibinfo {pages} {1903--1907} (\bibinfo
  {year} {2016})}\BibitemShut {NoStop}%
\bibitem [{\citenamefont {Greaney}\ \emph {et~al.}(2011)\citenamefont
  {Greaney}, \citenamefont {Lani}, \citenamefont {Cicero},\ and\ \citenamefont
  {Grossman}}]{greaney2011mpemba}%
  \BibitemOpen
  \bibfield  {author} {\bibinfo {author} {\bibfnamefont {P~Alex}\ \bibnamefont
  {Greaney}}, \bibinfo {author} {\bibfnamefont {Giovanna}\ \bibnamefont
  {Lani}}, \bibinfo {author} {\bibfnamefont {Giancarlo}\ \bibnamefont
  {Cicero}}, \ and\ \bibinfo {author} {\bibfnamefont {Jeffrey~C}\ \bibnamefont
  {Grossman}},\ }\bibfield  {title} {\enquote {\bibinfo {title} {Mpemba-like
  behavior in carbon nanotube resonators},}\ }\href@noop {} {\bibfield
  {journal} {\bibinfo  {journal} {Metallurgical and Materials Transactions A}\
  }\textbf {\bibinfo {volume} {42}},\ \bibinfo {pages} {3907--3912} (\bibinfo
  {year} {2011})}\BibitemShut {NoStop}%
\bibitem [{\citenamefont {Gal}\ and\ \citenamefont
  {Raz}(2020)}]{PhysRevLett.124.060602}%
  \BibitemOpen
  \bibfield  {author} {\bibinfo {author} {\bibfnamefont {A.}~\bibnamefont
  {Gal}}\ and\ \bibinfo {author} {\bibfnamefont {O.}~\bibnamefont {Raz}},\
  }\bibfield  {title} {\enquote {\bibinfo {title} {Precooling strategy allows
  exponentially faster heating},}\ }\href {\doibase
  10.1103/PhysRevLett.124.060602} {\bibfield  {journal} {\bibinfo  {journal}
  {Phys. Rev. Lett.}\ }\textbf {\bibinfo {volume} {124}},\ \bibinfo {pages}
  {060602} (\bibinfo {year} {2020})}\BibitemShut {NoStop}%
\bibitem [{\citenamefont {Ch{\'{e}}trite}\ \emph {et~al.}(2021)\citenamefont
  {Ch{\'{e}}trite}, \citenamefont {Kumar},\ and\ \citenamefont
  {Bechhoefer}}]{Ch_trite_2021}%
  \BibitemOpen
  \bibfield  {author} {\bibinfo {author} {\bibfnamefont {Raphaël}\
  \bibnamefont {Ch{\'{e}}trite}}, \bibinfo {author} {\bibfnamefont {Avinash}\
  \bibnamefont {Kumar}}, \ and\ \bibinfo {author} {\bibfnamefont {John}\
  \bibnamefont {Bechhoefer}},\ }\bibfield  {title} {\enquote {\bibinfo {title}
  {The metastable mpemba effect corresponds to a non-monotonic temperature
  dependence of extractable work},}\ }\href {\doibase 10.3389/fphy.2021.654271}
  {\bibfield  {journal} {\bibinfo  {journal} {Frontiers in Physics}\ }\textbf
  {\bibinfo {volume} {9}} (\bibinfo {year} {2021}),\
  10.3389/fphy.2021.654271}\BibitemShut {NoStop}%
\bibitem [{\citenamefont {Lin}\ \emph {et~al.}(2022)\citenamefont {Lin},
  \citenamefont {Li}, \citenamefont {He}, \citenamefont {Ren},\ and\
  \citenamefont {Wang}}]{PhysRevE.105.014104}%
  \BibitemOpen
  \bibfield  {author} {\bibinfo {author} {\bibfnamefont {Jie}\ \bibnamefont
  {Lin}}, \bibinfo {author} {\bibfnamefont {Kai}\ \bibnamefont {Li}}, \bibinfo
  {author} {\bibfnamefont {Jizhou}\ \bibnamefont {He}}, \bibinfo {author}
  {\bibfnamefont {Jie}\ \bibnamefont {Ren}}, \ and\ \bibinfo {author}
  {\bibfnamefont {Jianhui}\ \bibnamefont {Wang}},\ }\bibfield  {title}
  {\enquote {\bibinfo {title} {Power statistics of otto heat engines with the
  mpemba effect},}\ }\href {\doibase 10.1103/PhysRevE.105.014104} {\bibfield
  {journal} {\bibinfo  {journal} {Phys. Rev. E}\ }\textbf {\bibinfo {volume}
  {105}},\ \bibinfo {pages} {014104} (\bibinfo {year} {2022})}\BibitemShut
  {NoStop}%
\bibitem [{\citenamefont {Militaru}\ \emph {et~al.}(2021)\citenamefont
  {Militaru}, \citenamefont {Lasanta}, \citenamefont {Frimmer}, \citenamefont
  {Bonilla}, \citenamefont {Novotny},\ and\ \citenamefont
  {Rica}}]{memoryeffect1}%
  \BibitemOpen
  \bibfield  {author} {\bibinfo {author} {\bibfnamefont {Andrei}\ \bibnamefont
  {Militaru}}, \bibinfo {author} {\bibfnamefont {Antonio}\ \bibnamefont
  {Lasanta}}, \bibinfo {author} {\bibfnamefont {Martin}\ \bibnamefont
  {Frimmer}}, \bibinfo {author} {\bibfnamefont {Luis}\ \bibnamefont {Bonilla}},
  \bibinfo {author} {\bibfnamefont {Lukas}\ \bibnamefont {Novotny}}, \ and\
  \bibinfo {author} {\bibfnamefont {Raúl}\ \bibnamefont {Rica}},\ }\bibfield
  {title} {\enquote {\bibinfo {title} {Kovacs memory effect with an optically
  levitated nanoparticle},}\ }\href {\doibase 10.1103/PhysRevLett.127.130603}
  {\bibfield  {journal} {\bibinfo  {journal} {Physical Review Letters}\
  }\textbf {\bibinfo {volume} {127}} (\bibinfo {year} {2021}),\
  10.1103/PhysRevLett.127.130603}\BibitemShut {NoStop}%
\bibitem [{\citenamefont {Mompó}\ \emph {et~al.}(2021)\citenamefont {Mompó},
  \citenamefont {López-Castaño}, \citenamefont {Lasanta}, \citenamefont
  {Reyes},\ and\ \citenamefont {Torrente}}]{memoryeffect2}%
  \BibitemOpen
  \bibfield  {author} {\bibinfo {author} {\bibfnamefont {E.}~\bibnamefont
  {Mompó}}, \bibinfo {author} {\bibfnamefont {Miguel~Ángel}\ \bibnamefont
  {López-Castaño}}, \bibinfo {author} {\bibfnamefont {Antonio}\ \bibnamefont
  {Lasanta}}, \bibinfo {author} {\bibfnamefont {Francisco}\ \bibnamefont
  {Reyes}}, \ and\ \bibinfo {author} {\bibfnamefont {Aurora}\ \bibnamefont
  {Torrente}},\ }\bibfield  {title} {\enquote {\bibinfo {title} {Memory effects
  in a gas of viscoelastic particles},}\ }\href {\doibase 10.1063/5.0050804}
  {\bibfield  {journal} {\bibinfo  {journal} {Physics of Fluids}\ }\textbf
  {\bibinfo {volume} {33}},\ \bibinfo {pages} {062005} (\bibinfo {year}
  {2021})}\BibitemShut {NoStop}%
\bibitem [{\citenamefont {Gonz\'alez-Adalid~Pemart\'{\i}n}\ \emph
  {et~al.}(2021)\citenamefont {Gonz\'alez-Adalid~Pemart\'{\i}n}, \citenamefont
  {Momp\'o}, \citenamefont {Lasanta}, \citenamefont {Mart\'{\i}n-Mayor},\ and\
  \citenamefont {Salas}}]{memoryeffect3}%
  \BibitemOpen
  \bibfield  {author} {\bibinfo {author} {\bibfnamefont {Isidoro}\ \bibnamefont
  {Gonz\'alez-Adalid~Pemart\'{\i}n}}, \bibinfo {author} {\bibfnamefont
  {Emanuel}\ \bibnamefont {Momp\'o}}, \bibinfo {author} {\bibfnamefont
  {Antonio}\ \bibnamefont {Lasanta}}, \bibinfo {author} {\bibfnamefont
  {V\'{\i}ctor}\ \bibnamefont {Mart\'{\i}n-Mayor}}, \ and\ \bibinfo {author}
  {\bibfnamefont {Jes\'us}\ \bibnamefont {Salas}},\ }\bibfield  {title}
  {\enquote {\bibinfo {title} {Slow growth of magnetic domains helps fast
  evolution routes for out-of-equilibrium dynamics},}\ }\href {\doibase
  10.1103/PhysRevE.104.044114} {\bibfield  {journal} {\bibinfo  {journal}
  {Phys. Rev. E}\ }\textbf {\bibinfo {volume} {104}},\ \bibinfo {pages}
  {044114} (\bibinfo {year} {2021})}\BibitemShut {NoStop}%
\bibitem [{\citenamefont {Patr{\'{o} }n}\ \emph {et~al.}(2021)\citenamefont
  {Patr{\'{o} }n}, \citenamefont {S{\'{a}}nchez-Rey},\ and\ \citenamefont
  {Prados}}]{memoryeffect4}%
  \BibitemOpen
  \bibfield  {author} {\bibinfo {author} {\bibfnamefont {A.}~\bibnamefont
  {Patr{\'{o} }n}}, \bibinfo {author} {\bibfnamefont {B.}~\bibnamefont
  {S{\'{a}}nchez-Rey}}, \ and\ \bibinfo {author} {\bibfnamefont
  {A.}~\bibnamefont {Prados}},\ }\bibfield  {title} {\enquote {\bibinfo {title}
  {Strong nonexponential relaxation and memory effects in a fluid with
  nonlinear drag},}\ }\href {\doibase 10.1103/physreve.104.064127} {\bibfield
  {journal} {\bibinfo  {journal} {Physical Review E}\ }\textbf {\bibinfo
  {volume} {104}} (\bibinfo {year} {2021}),\
  10.1103/physreve.104.064127}\BibitemShut {NoStop}%
\bibitem [{\citenamefont {Uskokovic}(2020)}]{memoryeffect5}%
  \BibitemOpen
  \bibfield  {author} {\bibinfo {author} {\bibfnamefont {Vuk}\ \bibnamefont
  {Uskokovic}},\ }\bibfield  {title} {\enquote {\bibinfo {title} {…and all
  the world a dream: Memory effects outlining the path to explaining the
  strange temperature-dependency of crystallization of water, a.k.a. the mpemba
  effect},}\ }\href {\doibase 10.13128/Substantia-895} {\ \textbf {\bibinfo
  {volume} {4}},\ \bibinfo {pages} {59 -- 117} (\bibinfo {year}
  {2020})}\BibitemShut {NoStop}%
\bibitem [{\citenamefont {Teza}\ \emph
  {et~al.}(2023{\natexlab{b}})\citenamefont {Teza}, \citenamefont {Yaacoby},\
  and\ \citenamefont {Raz}}]{Teza2022-Eigenvalue}%
  \BibitemOpen
  \bibfield  {author} {\bibinfo {author} {\bibfnamefont {Gianluca}\
  \bibnamefont {Teza}}, \bibinfo {author} {\bibfnamefont {Ran}\ \bibnamefont
  {Yaacoby}}, \ and\ \bibinfo {author} {\bibfnamefont {Oren}\ \bibnamefont
  {Raz}},\ }\bibfield  {title} {\enquote {\bibinfo {title} {Eigenvalue crossing
  as a phase transition in relaxation dynamics},}\ }\href {\doibase
  10.1103/PhysRevLett.130.207103} {\bibfield  {journal} {\bibinfo  {journal}
  {Phys. Rev. Lett.}\ }\textbf {\bibinfo {volume} {130}},\ \bibinfo {pages}
  {207103} (\bibinfo {year} {2023}{\natexlab{b}})}\BibitemShut {NoStop}%
\bibitem [{\citenamefont {Walker}\ \emph {et~al.}(2020)\citenamefont {Walker},
  \citenamefont {Tam},\ and\ \citenamefont {Jarrell}}]{Walker_2020}%
  \BibitemOpen
  \bibfield  {author} {\bibinfo {author} {\bibfnamefont {Nicholas}\
  \bibnamefont {Walker}}, \bibinfo {author} {\bibfnamefont {Ka-Ming}\
  \bibnamefont {Tam}}, \ and\ \bibinfo {author} {\bibfnamefont {Mark}\
  \bibnamefont {Jarrell}},\ }\bibfield  {title} {\enquote {\bibinfo {title}
  {Deep learning on the 2-dimensional ising model to extract the crossover
  region with a variational autoencoder},}\ }\href {\doibase
  10.1038/s41598-020-69848-5} {\bibfield  {journal} {\bibinfo  {journal}
  {Scientific Reports}\ }\textbf {\bibinfo {volume} {10}} (\bibinfo {year}
  {2020}),\ 10.1038/s41598-020-69848-5}\BibitemShut {NoStop}%
\bibitem [{\citenamefont {Morningstar}\ and\ \citenamefont
  {Melko}(2017)}]{https://doi.org/10.48550/arxiv.1708.04622}%
  \BibitemOpen
  \bibfield  {author} {\bibinfo {author} {\bibfnamefont {Alan}\ \bibnamefont
  {Morningstar}}\ and\ \bibinfo {author} {\bibfnamefont {Roger}\ \bibnamefont
  {Melko}},\ }\bibfield  {title} {\enquote {\bibinfo {title} {Deep learning the
  ising model near criticality},}\ }\href@noop {} {\bibfield  {journal}
  {\bibinfo  {journal} {Journal of Machine Learning Research}\ }\textbf
  {\bibinfo {volume} {18}} (\bibinfo {year} {2017})}\BibitemShut {NoStop}%
\bibitem [{\citenamefont {Portman}\ and\ \citenamefont
  {Tamblyn}(2017)}]{Portman_2017}%
  \BibitemOpen
  \bibfield  {author} {\bibinfo {author} {\bibfnamefont {Nataliya}\
  \bibnamefont {Portman}}\ and\ \bibinfo {author} {\bibfnamefont {Isaac}\
  \bibnamefont {Tamblyn}},\ }\bibfield  {title} {\enquote {\bibinfo {title}
  {Sampling algorithms for validation of supervised learning models for
  ising-like systems},}\ }\href {\doibase 10.1016/j.jcp.2017.06.045} {\bibfield
   {journal} {\bibinfo  {journal} {Journal of Computational Physics}\ }\textbf
  {\bibinfo {volume} {350}},\ \bibinfo {pages} {871--890} (\bibinfo {year}
  {2017})}\BibitemShut {NoStop}%
\bibitem [{\citenamefont {Teza}\ and\ \citenamefont
  {Stella}(2020{\natexlab{a}})}]{Teza2021.125.110601}%
  \BibitemOpen
  \bibfield  {author} {\bibinfo {author} {\bibfnamefont {Gianluca}\
  \bibnamefont {Teza}}\ and\ \bibinfo {author} {\bibfnamefont {Attilio~L.}\
  \bibnamefont {Stella}},\ }\bibfield  {title} {\enquote {\bibinfo {title}
  {Exact coarse graining preserves entropy production out of equilibrium},}\
  }\href {\doibase 10.1103/PhysRevLett.125.110601} {\bibfield  {journal}
  {\bibinfo  {journal} {Phys. Rev. Lett.}\ }\textbf {\bibinfo {volume} {125}},\
  \bibinfo {pages} {110601} (\bibinfo {year} {2020}{\natexlab{a}})}\BibitemShut
  {NoStop}%
\bibitem [{\citenamefont {Teza}(2020)}]{Teza_Thesis}%
  \BibitemOpen
  \bibfield  {author} {\bibinfo {author} {\bibfnamefont {Gianluca}\
  \bibnamefont {Teza}},\ }\href@noop {} {\enquote {\bibinfo {title} {Out of
  equilibrium dynamics: from an entropy of the growth to the growth of entropy
  production.}}\ } (\bibinfo {year} {2020})\BibitemShut {NoStop}%
\bibitem [{\citenamefont {Teza}\ and\ \citenamefont
  {Stella}(2020{\natexlab{b}})}]{Teza.125.110601}%
  \BibitemOpen
  \bibfield  {author} {\bibinfo {author} {\bibfnamefont {Gianluca}\
  \bibnamefont {Teza}}\ and\ \bibinfo {author} {\bibfnamefont {Attilio~L.}\
  \bibnamefont {Stella}},\ }\bibfield  {title} {\enquote {\bibinfo {title}
  {Exact coarse graining preserves entropy production out of equilibrium},}\
  }\href {\doibase 10.1103/PhysRevLett.125.110601} {\bibfield  {journal}
  {\bibinfo  {journal} {Phys. Rev. Lett.}\ }\textbf {\bibinfo {volume} {125}},\
  \bibinfo {pages} {110601} (\bibinfo {year} {2020}{\natexlab{b}})}\BibitemShut
  {NoStop}%
\bibitem [{\citenamefont {Teza}\ \emph {et~al.}(2022)\citenamefont {Teza},
  \citenamefont {Yaacoby},\ and\ \citenamefont {Raz}}]{Teza2022.2203.11644}%
  \BibitemOpen
  \bibfield  {author} {\bibinfo {author} {\bibfnamefont {Gianluca}\
  \bibnamefont {Teza}}, \bibinfo {author} {\bibfnamefont {Ran}\ \bibnamefont
  {Yaacoby}}, \ and\ \bibinfo {author} {\bibfnamefont {Oren}\ \bibnamefont
  {Raz}},\ }\href {\doibase 10.48550/ARXIV.2203.11644} {\enquote {\bibinfo
  {title} {Far from equilibrium relaxation in the weak coupling limit},}\ }
  (\bibinfo {year} {2022})\BibitemShut {NoStop}%
\bibitem [{\citenamefont {Arnoldi}(1951)}]{arnoldi:principle}%
  \BibitemOpen
  \bibfield  {author} {\bibinfo {author} {\bibfnamefont {Walter~Edwin}\
  \bibnamefont {Arnoldi}},\ }\bibfield  {title} {\enquote {\bibinfo {title}
  {The principle of minimized iterations in the solution of the matrix
  eigenvalue problem},}\ }\href@noop {} {\bibfield  {journal} {\bibinfo
  {journal} {Quarterly of applied mathematics}\ }\textbf {\bibinfo {volume}
  {9}},\ \bibinfo {pages} {17--29} (\bibinfo {year} {1951})}\BibitemShut
  {NoStop}%
\bibitem [{\citenamefont {Breiman}\ \emph {et~al.}(1984)\citenamefont
  {Breiman}, \citenamefont {Friedman}, \citenamefont {Olshen},\ and\
  \citenamefont {Stone}}]{breiman:book84}%
  \BibitemOpen
  \bibfield  {author} {\bibinfo {author} {\bibfnamefont {Leo}\ \bibnamefont
  {Breiman}}, \bibinfo {author} {\bibfnamefont {Jerome~H.}\ \bibnamefont
  {Friedman}}, \bibinfo {author} {\bibfnamefont {Richard}\ \bibnamefont
  {Olshen}}, \ and\ \bibinfo {author} {\bibfnamefont {Charles}\ \bibnamefont
  {Stone}},\ }\href@noop {} {\emph {\bibinfo {title} {Classification and
  Regression Trees}}},\ CART\ (\bibinfo  {publisher} {Wadsworth and Brooks},\
  \bibinfo {address} {Monterey, CA},\ \bibinfo {year} {1984})\BibitemShut
  {NoStop}%
\bibitem [{\citenamefont {Hodson}\ \emph {et~al.}(2021)\citenamefont {Hodson},
  \citenamefont {Over},\ and\ \citenamefont
  {Foks}}]{https://doi.org/10.1029/2021MS002681}%
  \BibitemOpen
  \bibfield  {author} {\bibinfo {author} {\bibfnamefont {Timothy~O.}\
  \bibnamefont {Hodson}}, \bibinfo {author} {\bibfnamefont {Thomas~M.}\
  \bibnamefont {Over}}, \ and\ \bibinfo {author} {\bibfnamefont {Sydney~S.}\
  \bibnamefont {Foks}},\ }\bibfield  {title} {\enquote {\bibinfo {title} {Mean
  squared error, deconstructed},}\ }\href {\doibase
  https://doi.org/10.1029/2021MS002681} {\bibfield  {journal} {\bibinfo
  {journal} {Journal of Advances in Modeling Earth Systems}\ }\textbf {\bibinfo
  {volume} {13}},\ \bibinfo {pages} {e2021MS002681} (\bibinfo {year} {2021})},\
  \bibinfo {note} {e2021MS002681 2021MS002681}\BibitemShut {NoStop}%
\bibitem [{\citenamefont {Higgins}(2020)}]{Legendre}%
  \BibitemOpen
  \bibfield  {author} {\bibinfo {author} {\bibfnamefont {Brian}\ \bibnamefont
  {Higgins}},\ }\bibfield  {title} {\enquote {\bibinfo {title} {Fitting data
  with legendre polynomials},}\ }\href@noop {} {\  (\bibinfo {year}
  {2020})}\BibitemShut {NoStop}%
\bibitem [{\citenamefont {Tibshirani}(1996)}]{10.2307/2346178}%
  \BibitemOpen
  \bibfield  {author} {\bibinfo {author} {\bibfnamefont {Robert}\ \bibnamefont
  {Tibshirani}},\ }\bibfield  {title} {\enquote {\bibinfo {title} {Regression
  shrinkage and selection via the lasso},}\ }\href
  {http://www.jstor.org/stable/2346178} {\bibfield  {journal} {\bibinfo
  {journal} {Journal of the Royal Statistical Society. Series B
  (Methodological)}\ }\textbf {\bibinfo {volume} {58}},\ \bibinfo {pages}
  {267--288} (\bibinfo {year} {1996})}\BibitemShut {NoStop}%
\bibitem [{\citenamefont {Guegan}\ \emph {et~al.}(2015)\citenamefont {Guegan},
  \citenamefont {Hardin}, \citenamefont {Stevens},\ and\ \citenamefont
  {Williams}}]{Guegan:2015mea}%
  \BibitemOpen
  \bibfield  {author} {\bibinfo {author} {\bibfnamefont {Baptiste}\
  \bibnamefont {Guegan}}, \bibinfo {author} {\bibfnamefont {John}\ \bibnamefont
  {Hardin}}, \bibinfo {author} {\bibfnamefont {Justin}\ \bibnamefont
  {Stevens}}, \ and\ \bibinfo {author} {\bibfnamefont {Mike}\ \bibnamefont
  {Williams}},\ }\bibfield  {title} {\enquote {\bibinfo {title} {{Model
  selection for amplitude analysis}},}\ }\href {\doibase
  10.1088/1748-0221/10/09/P09002} {\bibfield  {journal} {\bibinfo  {journal}
  {JINST}\ }\textbf {\bibinfo {volume} {10}},\ \bibinfo {pages} {P09002}
  (\bibinfo {year} {2015})},\ \Eprint {http://arxiv.org/abs/1505.05133}
  {arXiv:1505.05133 [physics.data-an]} \BibitemShut {NoStop}%
\bibitem [{\citenamefont {Sadasivan}\ \emph {et~al.}(2019)\citenamefont
  {Sadasivan}, \citenamefont {Mai},\ and\ \citenamefont
  {D\"oring}}]{Sadasivan:2018jig}%
  \BibitemOpen
  \bibfield  {author} {\bibinfo {author} {\bibfnamefont {D.}~\bibnamefont
  {Sadasivan}}, \bibinfo {author} {\bibfnamefont {M.}~\bibnamefont {Mai}}, \
  and\ \bibinfo {author} {\bibfnamefont {M.}~\bibnamefont {D\"oring}},\
  }\bibfield  {title} {\enquote {\bibinfo {title} {{S- and p-wave structure of
  $S=-1$ meson-baryon scattering in the resonance region}},}\ }\href {\doibase
  10.1016/j.physletb.2018.12.035} {\bibfield  {journal} {\bibinfo  {journal}
  {Phys. Lett. B}\ }\textbf {\bibinfo {volume} {789}},\ \bibinfo {pages}
  {329--335} (\bibinfo {year} {2019})},\ \Eprint
  {http://arxiv.org/abs/1805.04534} {arXiv:1805.04534 [nucl-th]} \BibitemShut
  {NoStop}%
\bibitem [{\citenamefont {Landay}\ \emph {et~al.}(2019)\citenamefont {Landay},
  \citenamefont {Mai}, \citenamefont {D\"oring}, \citenamefont {Haberzettl},\
  and\ \citenamefont {Nakayama}}]{Landay:2018wgf}%
  \BibitemOpen
  \bibfield  {author} {\bibinfo {author} {\bibfnamefont {J.}~\bibnamefont
  {Landay}}, \bibinfo {author} {\bibfnamefont {M.}~\bibnamefont {Mai}},
  \bibinfo {author} {\bibfnamefont {M.}~\bibnamefont {D\"oring}}, \bibinfo
  {author} {\bibfnamefont {H.}~\bibnamefont {Haberzettl}}, \ and\ \bibinfo
  {author} {\bibfnamefont {K.}~\bibnamefont {Nakayama}},\ }\bibfield  {title}
  {\enquote {\bibinfo {title} {{Towards the Minimal Spectrum of Excited
  Baryons}},}\ }\href {\doibase 10.1103/PhysRevD.99.016001} {\bibfield
  {journal} {\bibinfo  {journal} {Phys. Rev. D}\ }\textbf {\bibinfo {volume}
  {99}},\ \bibinfo {pages} {016001} (\bibinfo {year} {2019})},\ \Eprint
  {http://arxiv.org/abs/1810.00075} {arXiv:1810.00075 [nucl-th]} \BibitemShut
  {NoStop}%
\bibitem [{\citenamefont {Xu}\ \emph {et~al.}(2022)\citenamefont {Xu},
  \citenamefont {Lin},\ and\ \citenamefont {Gowen}}]{XU2022100175}%
  \BibitemOpen
  \bibfield  {author} {\bibinfo {author} {\bibfnamefont {Jun-Li}\ \bibnamefont
  {Xu}}, \bibinfo {author} {\bibfnamefont {Xiaohui}\ \bibnamefont {Lin}}, \
  and\ \bibinfo {author} {\bibfnamefont {Aoife~A.}\ \bibnamefont {Gowen}},\
  }\bibfield  {title} {\enquote {\bibinfo {title} {Combining machine learning
  with meta-analysis for predicting cytotoxicity of micro- and nanoplastics},}\
  }\href {\doibase https://doi.org/10.1016/j.hazadv.2022.100175} {\bibfield
  {journal} {\bibinfo  {journal} {Journal of Hazardous Materials Advances}\
  }\textbf {\bibinfo {volume} {8}},\ \bibinfo {pages} {100175} (\bibinfo {year}
  {2022})}\BibitemShut {NoStop}%
\bibitem [{\citenamefont {Kraljevski}\ \emph {et~al.}(2023)\citenamefont
  {Kraljevski}, \citenamefont {Duckhorn}, \citenamefont {Tschöpe},
  \citenamefont {Schubert},\ and\ \citenamefont {Wolff}}]{2023}%
  \BibitemOpen
  \bibfield  {author} {\bibinfo {author} {\bibfnamefont {Ivan}\ \bibnamefont
  {Kraljevski}}, \bibinfo {author} {\bibfnamefont {Frank}\ \bibnamefont
  {Duckhorn}}, \bibinfo {author} {\bibfnamefont {Constanze}\ \bibnamefont
  {Tschöpe}}, \bibinfo {author} {\bibfnamefont {Frank}\ \bibnamefont
  {Schubert}}, \ and\ \bibinfo {author} {\bibfnamefont {Matthias}\ \bibnamefont
  {Wolff}},\ }\bibfield  {title} {\enquote {\bibinfo {title} {Paper tissue
  softness rating by acoustic emission analysis},}\ }\href {\doibase
  10.3390/app13031670} {\bibfield  {journal} {\bibinfo  {journal} {Applied
  Sciences}\ }\textbf {\bibinfo {volume} {13}},\ \bibinfo {pages} {1670}
  (\bibinfo {year} {2023})}\BibitemShut {NoStop}%
\bibitem [{\citenamefont {Press}\ \emph {et~al.}(1992)\citenamefont {Press},
  \citenamefont {Teukolsky}, \citenamefont {Vetterling},\ and\ \citenamefont
  {Flannery}}]{PresTeukVettFlan92}%
  \BibitemOpen
  \bibfield  {author} {\bibinfo {author} {\bibfnamefont {William~H.}\
  \bibnamefont {Press}}, \bibinfo {author} {\bibfnamefont {Saul~A.}\
  \bibnamefont {Teukolsky}}, \bibinfo {author} {\bibfnamefont {William~T.}\
  \bibnamefont {Vetterling}}, \ and\ \bibinfo {author} {\bibfnamefont
  {Brian~P.}\ \bibnamefont {Flannery}},\ }\href@noop {} {\emph {\bibinfo
  {title} {Numerical Recipes in C}}},\ \bibinfo {edition} {2nd}\ ed.\ (\bibinfo
   {publisher} {Cambridge University Press},\ \bibinfo {address} {Cambridge,
  USA},\ \bibinfo {year} {1992})\BibitemShut {NoStop}%
\end{thebibliography}%

\begin{onecolumngrid}

\appendix

\section{Additional Data}
In this section we give data for reference that is not used in the main analysis and discussion of the paper. Tab.~\ref{tab:extrapolation} gives the extrapolation accuracy of the methods. Positive accuracies of this table are plotted in Fig.~\ref{fig:PositiveExtrapolation} and the format is equivalent to Tab.~\ref{tab:results}.
\begin{table}[h]
\begin{tabular}{ |p{2.1cm}||p{3.5cm}|p{3.5cm}|p{3.5cm}|p{3.5cm}| }
 \hline
Excluded Values of $N$ & DT & NN & NN2  &  NLR\\
\hline

  \hline
  
   & & \textbf{Positive Accuracy} & &  \\
  
 \hline
 \hline
 
 15  &  0.5436 $\pm$ 0.018& 0.6638 $\pm$ 0.125  & 0.8421 $\pm$ 0.0724 & 0.3435 $\pm$ 0.03047\\
 13,15  & 0.5334 $\pm$ 0.0184  & 0.77960$\pm$0.0633 & 0.8040 $\pm$ 0.04347 & 0.25629 $\pm$ 0.06267\\
  11,13,15  & 0.5479 $\pm$ 0.0503 & 0.5922 $\pm$ 0.1994  & 0.6611$\pm$ 0.04952 & 0.3467 $\pm$ 0.05206\\
  
    9,11,13,15  & 0.5155 $\pm$ 0.04509 & 0.3826 $\pm$ 0.2561 & 0.6472 $\pm$ 0.1640 & 0.1909 $\pm$ 0.01386\\
7,9,11,13,15  &  0.5283 $\pm$ 0.0356& 0.2741 $\pm$ 0.08402  & 0.5222 $\pm$ 0.1288 & 0.2184 $\pm$ 0.009012\\
 
 \hline
 
   \hline
  
   & & \textbf{Negative Accuracy} & & \\
  
 \hline
 \hline
 
 15  &  0.996207 $\pm$ 0.00245& 0.99723 $\pm$ 
 0.000954173 & 0.997952 $\pm$ 0.00042289 &  0.718728 $\pm$ 0.0392236\\
13,15 & 0.99757 $\pm$ 0.001162 & 0.992953 $\pm$ 0.00388419  & 0.996802 $\pm$ 0.00108171 & 0.83444 $\pm$ 0.04182\\
  11,13,15  &  0.998175 $\pm$ 0.000508  & 0.987025 $\pm$ 0.00648745 & 0.997013 $\pm$ 0.000912489 & 0.890572 $\pm$ 0.0353344\\
9,11,13,15  & 0.996978 $\pm$ 0.001237  & 0.958839 $\pm$ 0.0328791  & 0.996679 $\pm$ 0.00173007 &  0.837763 $\pm$ 0.034161\\

7,9,11,13,15  &  0.99767 $\pm$ 0.00033 & 0.960547 $\pm$ 0.0529037  & 0.992296 $\pm$ 0.01823 & 0.93159 $\pm$ 0.00735792\\

  \hline
\end{tabular}
\caption{Extrapolation accuracy. The data is trained on 50,000 data points. 
    }
    \label{tab:extrapolation}
\end{table}

Tab.~\ref{tab:lasso} gives the best fit parameters for non-linear regression. Parameters not included in the table are either mathematically identical to an excluded parameter or shown by the LASSO method to have less impact. The fit was performed with 50000 data points and corresponds to the fifth row of each section of Tab.~\ref{tab:results} for NLR, however, NLR for any amount of data points greater than 1000 had similar minima. Note that the fit was done to data with $|a_2|$ multiplied by a factor of 10000 as described in the text. To relate these results to data that has not been prepared in this way, one should divide each parameter by 10000.

\begin{table}
\begin{tabular}{|c|c|c|c|c|c|c|c|c|c|}
\hline
  $c_{0000}$   & 12.0493 & $c_{0001}$ & 56.5418
  & $c_{0002}$& -77.7456 &$c_{0003}$ &  -3.88612& $c_{0004}$ & 4.93836 \\
    \hline

      $c_{0011}$   & 11.9538 & $c_{0012}$ & -389.284
  & $c_{0013}$& -12.5517 &$c_{0014}$ &  -6.88152 &$c_{0023}$ & 4.51196\\
    \hline

\hline
  $c_{0024}$   & 0.37328 & $c_{0034}$ & 0.780776
  & $c_{0044}$& 67.0015 &$c_{0111}$ &  -11.4937& $c_{0112}$ & -77.7841 \\
    \hline

\hline
  $c_{0113}$   & -5.35089 & $c_{0114}$ & -1.60171
  & $c_{0221}$& -308.538 &$c_{0222}$ &  -11.4937& $c_{0224}$ & 87.8123 \\
    \hline

\hline
  $c_{0331}$   & -801.778 & $c_{0332}$ & 67.5452
  & $c_{0333}$& 458.608 &$c_{0334}$ &  183.366 & $c_{0441}$ & 335.435 \\
    \hline

\hline
  $c_{0442}$   & -418.124 & $c_{0443}$ & 37.4534
  & $c_{0444}$& -38.2862 &$c_{0123}$ &  -0.00109161 & $c_{0124}$ & 16.838 \\
    \hline

\hline
  $c_{0134}$   & 0.431855 & $c_{0234}$ & 0.00739725
  & $c_{1114}$& 0.294404 &$c_{3334}$ &  -16.9157 & $c_{4441}$ & -193.343\\
    \hline
\end{tabular}
\caption{LASSO parameters for a fit with 500000 data points. Fit parameters, $c_{k_1 k_2 k_3 k_4}$ are given in Eq.~\eqref{eq:lassofitfunction}. All parameters not present in this table are either excluded by the LASSO as described in the main text or unnecessary because they are multiplied by a term identical to that of one an included parameter.}
\label{tab:lasso}
\end{table}
\section{Neural Network Architecture Tests}
\label{sec:NeuralArchitecture}

In this work we employ a neural network with three hidden layers with a total of 75 nodes: 50 in the first layer, 20 in the second, and 5 in the third. In this Appendix we explain our choice of architecture.

There is no one unique best architecture for a neural network. However, if a neural network has too few nodes or too few layers, it will not be able to achieve a small error. If a neural network has too many nodes or too many layers it can take prohibitively long to train or find weights that are at a local minimum that is nowhere near the global minimum. Our approach is to vary both the number of nodes and the number of layers until we find the smallest possible network that cannot  significantly reduce the error by increasing the number of nodes or layers.

This variation is shown in Fig.~\ref{fig:nnplot}. Our process was to vary the number of layers and number of nodes as described in the caption. We used the adam optimizer with a training data and a validation data of 50000 data point. 200 epochs were employed with a batch size of 200. 

Above the chosen values denoted with the dashed line, there is a slight decrease in the error but it is very small. For values below the dashed line, there is a very large increase in the error.

Our choices are not the only possible choices. Slightly different values could also be chosen and these might lead to results that are slightly better, but these results give evidence that a significant improvement in the accuracy will not occur. On the other hand, if the system had been much smaller either in nodes or layers, the error and hence the predictions would be significantly worse. This shows that the chosen architecture is reasonable.

\begin{figure}[b]
\centerline{\includegraphics[scale=0.65]{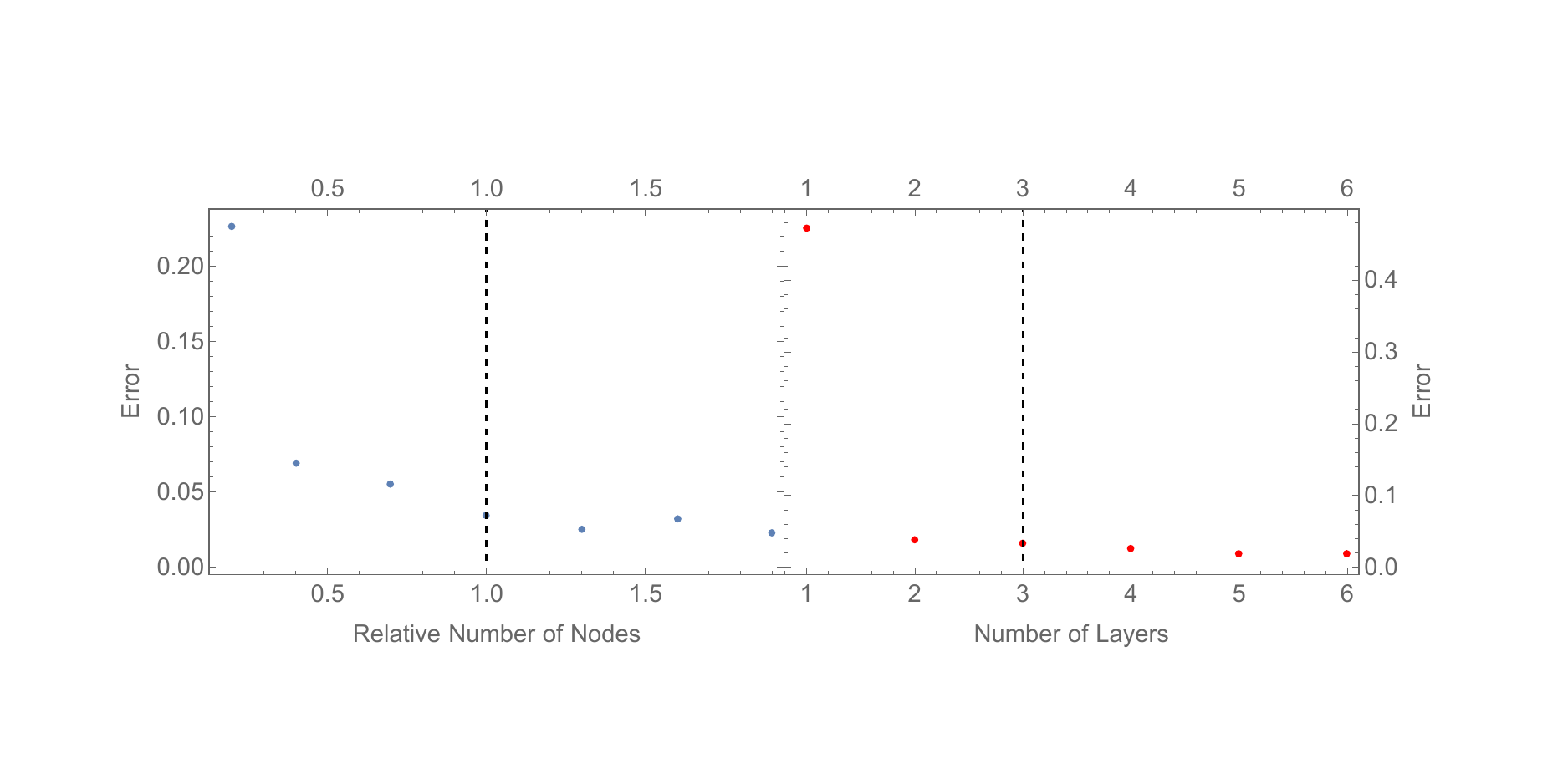}}
\caption{Plots of the accurace of the neural network (NN1) vs. quantities of the architecture. The left plot varies the relative number of nodes. In this plot there are three hidden layers of nodes. The ratios of each layer are held constant as the number of nodes is varied. The relative number is the number of nodes employed divided by our chosen number of hidden nodes, (75).  The right plot varies the number of layers. In this plot, the number of nodes is held constant. They are divided among the number of hidden layers on the horizontal axis. In both plots, the vertical dashed line indicates the values employed in this work.  }
\label{fig:nnplot}
\end{figure}

\end{onecolumngrid}


\end{document}